\documentclass[aps,amsmath,preprint,showpacs]{revtex4}
\usepackage{graphicx}
\usepackage{dcolumn}
\usepackage{color}

\newcommand{\tauiso}{{\mbox{\boldmath $\tau$}}}

\newcommand{\vecvarphi}{{\mbox{\boldmath $\varphi$}}}

\newcommand{\bm}{\bibitem}

\newcommand{\s}[1]{/\llap{$#1$}} 
\newcommand{\lrpart}{\tensor\partial}
\newcommand{\dop}[2]{#1\cdot#2}

\newcommand{\res}[2]{\ensuremath{#1_{#2}}}

\newcommand{\gamF}{\gamma_5}

\newcommand{\eps}{\varepsilon}

\newcommand{\A}{\ensuremath{\mathsf A}}

\newcommand{\K}{\ensuremath{\mathsf K}}
\newcommand{\T}{\ensuremath{\mathsf T}}
\newcommand{\tf}{\ensuremath{\tilde f}}

\newcommand{\La}{\ensuremath{\mathcal L}}

\newcommand{\oh}{\ensuremath{\frac12}}

\newcommand{\mc}[1]{\multicolumn{1}{c|}{#1}}
\newcommand{\mcc}[1]{\multicolumn{1}{c}{#1}}
\newcommand{\N}{\mc{---}}  
\newcommand{\Nc}{\mcc{---}}  

\newcommand{\ch}[4]{\ensuremath{#1 #2 \to #3 #4}}
\newcommand{\YYv}[1]
    {\frac{(\chi #1 + i \s\partial #1 /2m_N) \cdot \tauiso}
         {\chi+1} \gamF}
\newcommand{\YY}[1]
    {\frac{\chi #1 + i \s\partial #1 /2m_N}
         {\chi+1} \gamF}
\newcommand{\XXv}[1]
    {\left(\gamma_\mu \vecvarphi_\rho^\mu
    +\frac{\kappa_{\rho}}{2m_N} \sigma_{\mu\nu}\partial^\nu\vecvarphi_\rho^\mu
      \right) \cdot \tauiso}
\newcommand{\XX}[1]
    {\left(\gamma_\mu \varphi_\omega^\mu
    +\frac{\kappa_\omega}{2m_N} \sigma_{\mu\nu}\partial^\nu \varphi_\omega^\mu
      \right)}
\newcommand{\XXp}[1]
    {\left(\gamma_\mu \varphi_\phi^\mu
    +\frac{\kappa_\phi}{2m_N} \sigma_{\mu\nu}\partial^\nu \varphi_\phi^\mu
      \right)}
\newcommand{\EPS}[4]
    {\left(\eps_{\mu\nu\rho\sigma}
      (#1^\rho #2^\mu) (#3^\sigma #4^\nu) \right)}
\long\def\Omit#1{}

\newcolumntype{x}[1]{D..{#1}}
\newcolumntype{C}{c}
\newcolumntype{R}{r}

\newcommand{\tblref}[1]{Table~\ref{tbl:#1}}
\newcommand{\tbllab}[1]{\label{tbl:#1}}
\renewcommand{\eqref}[1]{Eq.~(\ref{eq:#1})}
\newcommand{\eqlab}[1]{\label{eq:#1}}

\newcommand{\secref}[1]{Section~\ref{sec:#1}}
\newcommand{\seclab}[1]{\label{sec:#1}}

\begin{document}

\title{The associated photoproduction of $K^+$ meson off proton 
within a coupled-channels \K-matrix approach}
\author{R. Shyam$^{1,3}$, O. Scholten$^2$, and H. Lenske$^1$}
\affiliation{$^1$Institute f\"ur Theoretische Physik, Universit\"at Giessen,
D-35392 Giessen, Germany \\ 
$^2$Kernfysisch Versneller Instituut, University of Groningen, 
9747 AA, Groningen, The Netherlands \\
$^3$Saha Institute of Nuclear Physics, Kolkata 70064,
India }

\date{\today}

\begin{abstract}

We investigate the $p(\gamma,K^+)\Lambda$ and $p(\gamma, K^+)\Sigma^0$
reactions within a coupled-channels effective-Lagrangian method which
is based on the K-matrix approach. The two-body final
channels included are $\pi N$, $\eta N$, $\phi N$, $\rho N$, $\gamma N$,
$K \Lambda$, and $K \Sigma$. Non-resonant meson-baryon interactions 
are included in the model via nucleon intermediate states in the $s$- 
and $u$-channels and meson exchanges in the $t$-channel amplitude and 
the $u$-channel resonances.  The nucleon resonances $S_{11}$(1535), 
$S_{11}$(1650), $S_{31}$(1620), $P_{11}$(1440), $P_{11}$(1710), 
P$_{13}$(1720), $P_{33}$(1232), $P_{33}$(1600), D$_{13}$(1520), 
$D_{13}$(1700), and $D_{33}$(1700) are included explicitly in the 
calculations. With a single parameter set which was derived earlier 
from our analysis of the $\eta$ meson photoproduction, the model 
describes well all the available cross section and polarization data 
of the SAPHIR collaboration for the two investigated channels. The 
description of the data of the CLAS collaboration, however, is not of 
the same quality. In contrast to some previous studies, we do not find 
any compelling need for including a $D_{13}$ state with mass of around 
2.0 GeV in order to reproduce the data for the $p(\gamma,K^+)\Lambda$
reaction at photon energies corresponding to the invariant mass 
around 1.9 GeV. 

\end{abstract}
\maketitle

\newpage
\section{Introduction}

Recently, several laboratories have devoted considerable amount of 
effort to the investigation of the photoproduction of strangeness 
on the nucleon. Experiments performed at JLab-CLAS~\cite{bra06,mcn04}, 
ELSA-SAPHIR~\cite{gla04,law05} and SPring8/LEPS~\cite{zeg03,sum06} 
have produced high quality data on the associate strangeness production 
reactions $p(\gamma,K^+)\Lambda$ and $p(\gamma, K^+)\Sigma^0$ covering 
the photon energy regime from threshold to upto 3.0 GeV. Furthermore, 
data on single and double polarization~\cite{bra07} observables have 
also become available.  The motivation behind these studies has been 
the fact that a considerable part of the excitation spectrum of the 
nucleon can, in principle, participate in the production process of 
associated strangeness ($K^+\Lambda$ and $\K^+\Sigma^0$) through these  
reactions because even at the threshold they involve invariant masses 
that exceed those of several baryonic resonances. It is hoped that 
these studies could help in investigating the so-called missing 
baryonic resonances that are predicted by the quark models (see, e.g., 
Ref.~\cite{cap00}) but not observed in non-strange meson photoproduction. 
Some of these resonances may couple strongly to the $K\Lambda$ and 
$K\Sigma$ channels.    

The determination of the properties of the nucleon resonances (e.g., 
their masses, widths, and coupling constants to various decay channels)
is an important issue in hadron physics. This will help in testing the 
predictions of lattice quantum chromodynamics (LQCD) which is the only 
theory that tries to calculate these properties from first principles. 
Even though, the requirement of computational power is enormous for 
their numerical realization, such calculations have started to provide 
results for the properties of nucleon ground as well as excited states.
\cite{and09,bas07,bur06,mat05,lei05}. Furthermore, reliable nucleon 
resonance data are important for testing the "quantum chromodynamics 
(QCD) based" quark models of the nucleon (see, {\it e.g.},
\cite{cap00,lor01}) and also the dynamical coupled-channels 
models of baryonic resonances~\cite{lut05}.

One of the major challenges of this field is the extraction of 
reliable information about nucleon resonance properties from the 
photoproduction data. In experiments, these resonances are excited 
as intermediate states before decaying into the final meson and baryon 
channels. A good description of intermediate-energy scattering 
is still not amenable to the LQCD calculations. Therefore, at this 
stage the effective methods are usually employed to describe the 
dynamics of the meson production reactions. Baryon resonance states 
are included explicitely in these approaches and their properties are 
obtained by comparing the predictions of the theory with the 
experimental data~\cite{ben95,arn00,tia99,dre99,shy07,jan01}.

In order to determine resonance properties reliably from the 
experimental measurements one requires a model that can analyze the 
different reactions over the entire energy range using a single 
Lagrangian density that generates all non-resonance contributions 
from Born, $u$- and $t$-channel contributions without introducing 
new parameters. At the same time, the Lagrangian should also satisfy 
the symmetries of the fundamental theory (i.e. QCD) while retaining 
only mesons and baryons as effective degrees of freedom.  

The coupled-channels method within the \K-matrix approach
\cite{feu98,kor98,pen02,uso05,shk05,uso06,shy08} provides a way 
to analyze simultaneously all reaction data for a multitude of 
observables in different reaction channels while respecting the 
constraints described above. This method is attractive because it is 
based on an effective-Lagrangian framework that is gauge invariant 
and is consistent with chiral symmetry. It also provides a convenient 
way of imposing the unitarity constraint. This results from the 
Bethe-Saltpeter equation in the approximation where  particles
forming the loop are taken on the mass shell i.e. only the 
discontinuity part of the loop integral is retained. The $S$ matrix 
in this approach is unitary provided the \K-matrix is taken to be 
real and Hermitian.

Alternatively, the dynamical coupled channels models within the 
Hamiltonian formalism have also been used to describe the 
meson-production reactions \cite{dur08,jul08,jul07,sag07,jul05,jul06,chi01}. 
Isobaric models such as Kaon-Maid~\cite{mar01a} and 
Saclay-Lyon~\cite{dav96} have been utilized in Ref.~\cite{mar01b,mar06} 
to describe the $p(\gamma,K^+)\Lambda$ reaction. This reaction has 
also been studied within a variety of tree-level isobar models
\cite{ade85,ade90,wil91,han01} and in the quark models~\cite{li95,lu95,he08}.
In Ref.~\cite{bor07} a gauge-invariant chiral unitary framework is
used to describe the photoproduction data of both SAPHIR and CLAS
collaborations.

The main objective of this paper is to study photoproduction
reactions $p(\gamma,K^+)\Lambda$ and $p(\gamma, K^+)\Sigma^0$ 
for photon energies ranging from threshold to about 3~GeV in a 
coupled-channels formalism of Refs.~\cite{uso05,uso06,shy08}
which is based on the \K-matrix approach. This is an effective 
Lagrangian model which is gauge invariant and obeys the low-energy 
theorem. We aim at describing simultaneously the data on total 
and differential cross sections as well as on polarization 
observables for both the reactions within the same framework with a 
single parameter set. The  $\Lambda$ and $\Sigma^0$ hyperons have 
isospins of 0 and 1, respectively. Thus, the intermediate states of 
$K^+\Lambda$ have the isospin $\frac{1}{2}$ only ($N^*$) while those 
of $K^+\Sigma^0$ can have both $\frac{1}{2}$ and $\frac{3}{2}$ 
isospins ($N^* $ and $\Delta$). Therefore, a combined description of 
all the available data for both these channels is indeed quite 
interesting and is a challenge to any theoretical model.  

We would like to add that a subset of SAPHIR and CLAS data for
the $p(\gamma,K^+)\Lambda$ reaction has been investigated previously 
in Ref.~\cite{shk05} within the Giessen model which is also a 
coupled-channels effective-Lagrangian \K-matrix approach. Despite some 
differences in details (see the discussions in Ref.~\cite{shy08}), our 
method is similar to that of the Giessen model. The calculations
presented in Ref.~\cite{shk05} have reproduced the $p(\gamma,K^+)\Lambda$
data well with a slight preference to the SAPHIR data. In the present
work we have attempted to describe the $p(\gamma,K^+)\Sigma^0$ reaction 
also along with the $p(\gamma,K^+)\Lambda$ one within the same framework 
using the same parameter set which already makes our analysis 
self-contained. Nevertheless, by comparing our results with those of
the Giessen model we expect to gain further insight in the mechanism
of the strangeness photoproduction off proton.  

Our paper is organized in the following way. An overview of our model 
is given in section II. This consists of a short discussion of the 
\K-matrix formalism, the model space and the channels included, the 
Lagrangians, and the form factors. Our results and their discussions 
are presented in section III. Summary and conclusions of our work are 
presented in section IV. Finally form of Lagrangians at various 
vertices are given in  appendix A.

\section{Description of the Model}

This work is based on an effective-Lagrangian model.  The kernel in the
\K-matrix approach is built by using the effective Lagrangian as 
given in Appendix A. We have taken into account contributions from (i)
the nucleon Born term, (ii) $t$-channel exchanges of mesons, (iii) 
nucleon and resonance terms in the $u$-channel, and (iv) baryonic 
resonance in the $s$-channel (see Fig.~1). The sum of amplitudes 
(i), (ii) and (iii) is termed as the background contribution. 
As is discussed below, this approach allows to account for 
coupled-channels effects while preserving many symmetries of a 
full field-theoretical method.
\begin{figure*}
\begin{center}
\includegraphics[width=0.5 \textwidth]{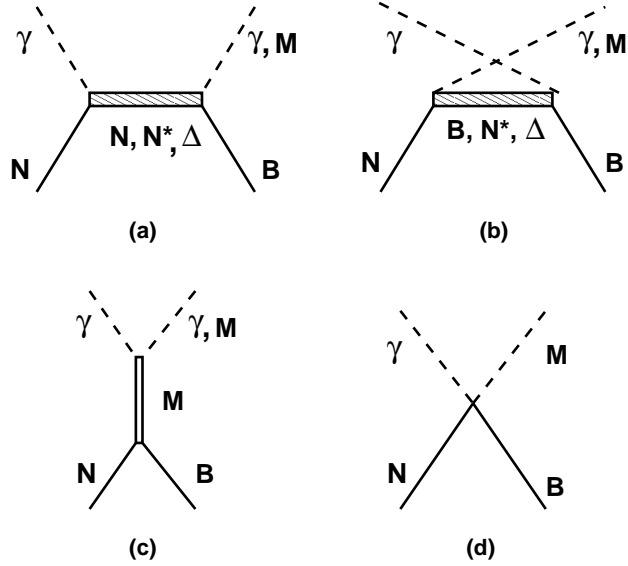}
\vskip -0.1in
\caption{[color online]
Feynman diagrams included in this work. First row: $s$- and $u$-channel
diagrams with propagating final state baryons ($B = N$, $\Lambda$, 
$\Sigma$) or intermediate state resonances ($\Delta$, $N^*$). $M$ 
stands for mesons included in the model space. Second row: 
$t$-channel contributions with propagating asymptotic and intermediate
mesons, and the contact term required by the gauge invariance.
}
\end{center}
\label{Fig1}
\end{figure*}

\subsection{\K-matrix model}\label{sec:Kmatr}

The coupled-channels (or re-scattering) effects are included in our 
model via the \K-matrix formalism. In this section we present a short 
overview of this approach; a more detailed description can be found 
in Refs.~\cite{kor98,uso05,sch02,new82}.

In the \K-matrix formalism the scattering matrix is written as
\begin{equation}\eqlab{T-matr}
  \T = \frac{\K}{1-i\K} \,.
\end{equation}

It is easy to check that the resulting scattering amplitude 
$S=1+2i\T$ is unitary provided that \K\ is Hermitian. The construction 
in \eqref{T-matr} can be regarded as the re-summation of an infinite 
series of loop diagrams by making a series expansion,
\begin{equation}
  \T = \K + i\K\K + i^2\K\K\K + \cdots \,.
\end{equation}
The product of two \K-matrices can be rewritten as a sum of different 
one-loop contributions (three- and four-point vertex and self-energy 
corrections) depending on Feynman diagrams that are included in the 
kernel \K. However, not the entire spectrum of loop corrections present 
in a true field-theoretical approach, is generated in this way and 
the missing ones should be accounted for in the kernel. In 
constructing the kernel, care should be taken to avoid double counting. 
For this reason we include in the kernel tree-level diagrams only 
[Figs. 1(a)-1(c)], modified with form-factors and contact terms 
[Fig. 1(d)]. The contact terms (or four-point vertices) ensure gauge 
invariance of the model and express model-dependence in working with 
form factors (see~\secref{ff}). Inclusion of both 
$s$- and $u$-channel diagrams [Figs 1(a) and 1(b), respectively] in 
the kernel insures the compliance with crossing symmetry.

{\squeezetable
\begin{table}
  \caption{ \tbllab{resonances}
    Baryon states included in the calculation of the
    kernel with their coupling constants. The column labeled
    WD lists the decay width to states outside the model space.
    The columns labeled M and WD are in units of GeV. See text
    for a discussion on the signs of the coupling constants.}
  \begin{ruledtabular}
  \begin{tabular}{C|x3|x3|d|d|d|d|d|d} 
    $L_{IJ}$ &\mc{M}&\mc{WD}&\mc{$g_{N\pi}$} &\mc{$g^1_{p\gamma}$}
    &\mc{$g^2_{p\gamma}$}&\mc{$g_{K\Lambda}$}&\mc{$g_{K\Sigma}$}
    &\mcc{$g_{N\eta}$} \\
    \hline
    \res{S}{11}(1535) &1.525 &0.0  &0.6  &-0.60& \N   &0.1   
&0.0 & 2.2 \\
    \res{S}{11}(1650) &1.690 &0.030&1.0  &-0.45& \N   &-0.1  
&0.0 & -0.8\\
    \res{S}{31}(1620) &1.630 &0.100&3.7  &-0.12& \N   & \N   
&-0.8&  \Nc\\
    \hline
    \res{P}{11}(1440) &1.520 &0.200&5.5  & 0.65& \N   &0.0   
&-2.0& 0.0 \\
    \res{P}{11}(1710) &1.850 &0.300&3.0  & 0.25& \N   &0.0   
&-3.0& 2.0 \\
    \res{P}{13}(1720) &1.750 &0.300&0.12 &-0.75&0.25  &-0.05 
&0.0 & 0.12 \\
    \res{P}{33}(1230) &1.230 &0.0  &1.7  &-2.2 &-2.7  & \N   
&0.0 & \Nc\\
    \res{P}{33}(1600) &1.855 &0.150&0.0  &-0.4 &-0.6  & \N   
&0.55& \Nc\\
    \hline
    \res{D}{13}(1520) &1.515 &0.050&1.2  & 2.6 & 2.5  &2.0   
&0.0 &1.2 \\
    \res{D}{13}(1700) &1.700 &0.090&0.0  &-0.5 & 0.0  &0.0   
&0.3 &-0.04\\
    \res{D}{33}(1700) &1.670 &0.250&0.8  & 1.5 & 0.6  & \N   
&-3.0& \Nc\\
  \end{tabular}
  \end{ruledtabular}
\end{table}
}

To be more specific, the loop corrections generated in the \K-matrix 
procedure include only those diagrams which correspond to two 
on-mass-shell particles in the loop~\cite{kon00,kon02}. This is the 
minimal set of diagrams one has to include to ensure two-particle 
unitarity. Thus, all diagrams that are not two particle reducible
(e.g. $\gamma N \to 2\pi N$ channels), are not included. In addition, 
only the convergent pole contributions i.e.\ the imaginary parts of 
the loop correction, are generated. The omitted real parts are 
important to guarantee analyticity of the amplitude and may have 
complicated cusp-like structures at energies where other reaction 
channels open. In principle, these can be included as form factors 
as is done in the dressed \K-matrix procedure~\cite{kon00,kor03}. 
We have chosen to work with purely phenomenological form factors   
for the reason of simplicity. An alternative procedure to account 
for the real-loop corrections is offered in Refs.
\cite{sat96,chi04,lut02}.

The strength of the \K-matrix procedure is that in spite of its 
simplicity, several symmetries are obeyed by it~\cite{sch02}. As 
was already noted, the resulting amplitude is unitary provided that 
\K\ is Hermitian, and it obeys gauge invariance provided the kernel 
is gauge-invariant. In addition, the scattering amplitude complies 
with crossing symmetry when the kernel is crossing symmetric. This 
property is crucial for a proper behavior of the scattering amplitude
in the low-energy limit~\cite{kon02,kon04}. Coupled-channels effects 
are automatically accounted for by this approach for the channels 
explicitly included into the \K-matrix as the final states.

As a result of this channel coupling, the resonances generate widths 
which are compatible with their decays to channels included in the 
model space. For some resonances, such as the $\Delta$ and the 
$S_{11}$(1535), this corresponds to their total width. Other resonances,
particularly the high lying ones, may have important decay branches to 
states that are not included in the model basis. To account for this 
in our calculations, we have added an explicit dissipative part to the 
corresponding propagators. The magnitudes of these widths are equivalent
to decay widths of the resonances to states outside of our model space.

The resonances which are taken into account in building the kernel 
are summarized in \tblref{resonances}. In the current work we limit 
ourselves to the spin-$\frac12$ and spin-$\frac32$ resonances as in 
this energy regime higher spin resonances are known~\cite{shk05} to 
give only a minor contribution to the $K^+\Lambda$ channel.  
Spin-$\frac32$ resonances are included with so-called 
gauge-invariant vertices which have the property that the coupling 
to spin-$\frac12$ pieces in the Rarita-Schwinger propagator vanish
\cite{pas00,kon00,pas01}. We have chosen this prescription since
it reduces the number of parameters as we do not have to deal with the
off-shell couplings. The effects of these off-shell couplings can be
absorbed in contact terms~\cite{pas01} which we prefer, certainly
within the context of the present work.  

The masses of the resonances given in \tblref{resonances} are bare 
masses and they thus may deviate from the values given by the 
Particle Data Group~\cite{PDG}. Higher-order effects in the \K-matrix 
formalism do give rise to a (small) shift of the pole-position with 
respect to the bare masses. The masses of very broad resonances, in 
particular the $P_{11}$, are not well determined - values lying in a 
broad range (typically a spread of the order of a quarter of the width)
give comparable results.  The width quoted in \tblref{resonances} 
corresponds to the partial width for decay to states outside our model 
space. The parameters as quoted in \tblref{resonances} are mostly 
unchanged as compared to those presented in previous calculations within
this model~\cite{uso05,uso06,shy08}. The $t$-channel contributions 
which are included in the kernel, are summarized in \tblref{particles}.
\begin{table}
  \caption{ \tbllab{particles}
    Mass, spin, parity and isospin of the mesons which are
    included in the model. The rightmost column 
    specifies in which reaction channels their $t$-channel
    contribution are taken into account. }
  \begin{ruledtabular}
  \begin{tabular}{C|d|CC|c}
    \mc{Meson}&\mc{M [GeV]}& S$^\pi$ & I & t-ch contributions \\
    \hline
    $\pi$     & 0.135      & 0$^-$ & 1 & (\ch \gamma N \phi N), (
\ch \pi N \rho N)\\
    K         & 0.494     & 0$^-$ &\oh& (\ch \gamma N K \Lambda), 
(\ch \gamma N K\Sigma) \\
    $\phi$    & 1.019     & 1$^-$ & 0 & \\
    $\eta$    & 0.547     & 0$^-$ & 0 & (\ch \gamma N \phi N) \\
    \hline
    $\rho$    & 0.770     & 1$^-$ & 1 & (\ch \gamma N \pi N), 
(\ch \gamma N \eta N), (\ch K\Lambda K\Sigma), \\ 
   &           &     &   & (\ch K\Sigma K\Sigma), (\ch N\pi K\Lambda),
 \\ 
   &           &     &   & (\ch N\pi N\eta), (\ch N\pi N\pi) \\
    $\omega$  & 0.781     & 1$^-$ & 0 & (\ch N\gamma N\pi),
        (\ch N\gamma N\eta) \\
    $\sigma$  & 0.760     & 0$^+$ & 0 & (\ch N\gamma N\phi), 
            (\ch N\pi N\pi) \\
    K$^*$   & 0.892     & 1$^-$ &\oh& (\ch N\gamma K\Lambda),
            (\ch N\gamma K\Sigma), \\
            &           &     &   & (\ch K\Lambda N\eta), 
(\ch K\Sigma N\eta), \\
            &           &     &   & (\ch N\pi K\Sigma) \\
  \end{tabular}
  \end{ruledtabular}
\end{table}

\subsection{Model space, channels included}

To keep the model manageable and relatively simple, we consider only 
stable particles or narrow resonances in two-body final states.  
The $\Lambda K$, $\Sigma K$, $N\phi$, $N\eta$ and $N\gamma$ are the 
final states of primary interest, and the $N\pi$ final state is 
included for its strong coupling to most of the resonances. Three-body 
final states, such as $2\pi N$, are not included explicitly as 
stated above. Their influence on widths of resonances is taken
into account by assigning an additional (energy dependent) width to 
them~\cite{kor98}. To investigate the effects of coupling to more
complicated states, we have also included the $N\rho$ final state. 
As was shown in Ref.~\cite{uso05}, inclusion of the $\rho$ channel has 
a strong influence on the pion sector but only a relatively minor 
effect on $\Lambda$ and $\Sigma$ photoproduction.

The components of the kernel which couple the different 
non-electromagnetic channels are taken as the sum of tree-level 
diagrams, similar to what is used for the photon channels. For these 
other channels no additional parameters were introduced and they 
thus need no further discussion.

\subsection{Form-factors \& gauge restoration \seclab{ff}}

Without the introduction of form factors, calculations with Born terms 
strongly overestimates the cross section at higher energies.
Although inclusion of coupled-channels effects reduces the cross 
section at high energies  yet disagreement with the experimental data 
still persists. Therefore, the Born contribution will have to be 
quenched with form factors. There are two physical motivations for 
introducing form factors (or vertex functions).  First of all, 
at high photon energies one may expect to become sensitive to the 
short-range quark structure of the nucleon. Because this physics is 
not included explicitly in our model, we can only account for it through
the introduction of phenomenological vertex functions. The second 
reason has to do with the intermediate-range effects because of 
meson-loop corrections which are not generated through the \K-matrix 
formalism. Examples of these are given in Refs.~\cite{kon00,kor03}.

In our approach as well as that of Ref.~\cite{pen02}, the form-factors 
are not known \emph{a priori} and thus they introduce certain 
arbitrariness in the model. In the current paper we limit ourselves 
to dipole form-factors in $s$-, $u$-, and $t$-channels because of 
their simplicity,
\begin{equation}\eqlab{ff-dipole}
  F_m(s)=\frac{\lambda^2}{\lambda^2+(s-m^2)^2} \;,
\end{equation}
where $m$ is the mass of the propagating particle and $\lambda$ is the
cut-off parameter. For ease of notation we introduce the subtracted form 
factors
\begin{equation}\eqlab{ff-twiddle}
  \tf_m(s)=\frac{1-F_m(s)}{s-m^2} \,,
\end{equation}
where $F_m(s)$ is normalized to unity on the mass-shell, 
$F_m(m^2)=1$, and $\tf_m(m^2)$ is finite.

However, only in the kaon sector we use a different functional form 
for the $u$-channel form-factors
\begin{equation}\eqlab{ff-u-channel}
  H_m(u)=\frac{u\lambda^2}{\big( \lambda^2+(u-m^2)^2 \big)m^2}.
\end{equation}
The argumentation for this different choice is presented in the 
discussion of the $\Sigma$-photoproduction results in Ref.~\cite{uso05}. 
Often a different functional form and cut-off values are introduced 
for the $t$-channel form factors. Although this can easily be motivated, 
it introduces additional model dependence and increases the number of 
free parameters. To limit the overall number of parameters we have 
taken the same cut-off value [$\lambda=1.2\; \mbox{GeV}^2$, see 
\eqref{ff-dipole}] for all form-factors except for the Born 
contributions in kaon channels where we used $\lambda=1.0\;\mbox{GeV}^2$.

Inclusion of form-factors will in general break electromagnetic
gauge-invariance of the model. Therefore, a gauge-restoration procedure 
should be applied. In Ref.~\cite{uso05}, the implications of various 
gauge-restoration procedures was studied for the $\gamma p \to K \Sigma$
amplitude. It was observed that the gauge-invariance restoration 
procedure is model dependent which may give rise to strongly different 
Born contributions to the amplitude. Therefore, the choice of a 
procedure to be adopted is guided by its ability to describe the 
experimental data.  It was found that the gauge-restoration procedure 
of Davidson and Workman~\cite{dav01} provided the best description of 
the data on the $K \Sigma$ photoproduction. We have used this procedure 
in the present work also.

We note that fitting the pion-scattering and pion-photoproduction 
amplitudes fixes masses as well as pion- and photon-coupling constants 
for most of the resonances. This limits strongly the number of free 
parameters for the kaon-production channels.

\section{Results and Discussions}

Our aim in this paper is to use the data base on 
$p(\gamma,K^+)\Lambda$ and $p(\gamma, K^+)\Sigma^0$ reactions of 
both SAPHIR and CLAS collaborations to check various ingredients 
of our unitary coupled-channels field theoretic model of meson 
production in photon induced reactions on nucleons. In particular we 
are interested in checking to what extent a simultaneous fit to data 
for a multitude of observables for both the reactions can be obtained 
with a single set of input parameters. This is expected to provide a 
strong constraint on the model parameters thus reducing the model 
dependence to a minimum. It is also likely to highlight the role
of channel couplings in various regions of photon energies because 
several calculations of the associated kaon production reactions
have neglected these effects.

We emphasize however, that even the large experimental 
data base may not allow to fix the extracted parameters uniquely 
within the unitary coupled-channels effective-Lagrangian model
\cite{pen02}. This is due to the fact that it is necessary to include 
empirical form factors in the model to regularize the amplitudes at 
higher energies. As mentioned before These form factors require a 
gauge-invariance restoration procedure which involves ambiguities. 
Nevertheless, confronting the model with a large data base in several 
reaction channels is expected to provide a means to overcome this 
problem.

The parameters in the model have been adjusted~\cite{uso06} to reproduce 
the Virginia Tech partial wave amplitudes of Arndt et al.~\cite{vir96}. 
In Ref.~\cite{shy08} we have presented a comparison of our calculated 
$S$-, $P$-, and $D$-wave amplitudes for pion-nucleon scattering for 
isospins $I$ = 1/2 and 3/2 channels with those of the FA08 single-energy 
partial wave amplitudes of Ref.~\cite{vir96}. The corresponding results 
for pion photoproduction and the Compton scattering are given in 
Ref.~\cite{uso06}. There we noted that both real and imaginary parts of 
the pion-nucleon scattering amplitudes are described well although some 
differences start to show up at the upper limit of the energy range 
considered.

The data for $p(\gamma,K^+)\Lambda$ and $p(\gamma, K^+)\Sigma^0$ 
reactions consist of total and differential cross sections and 
hyperon polarizations measured at CB-ELSA (SAPHIR collaboration 
\cite{gla04}) and at JLab (CLAS collaboration~\cite{mcn04,bra06}) 
for photon energies ranging from respective thresholds to about 
3~GeV. Moreover, beam asymmetry data are available for 9 photon 
energy beans between 1.5 GeV to 2.4 GeV from SPring8/LEPS
\cite{zeg03,sum06} group. These data, therefore, cover not only 
the entire resonance region but also the region where the background 
contributions are expected to be dominant. 
\begin{figure*}
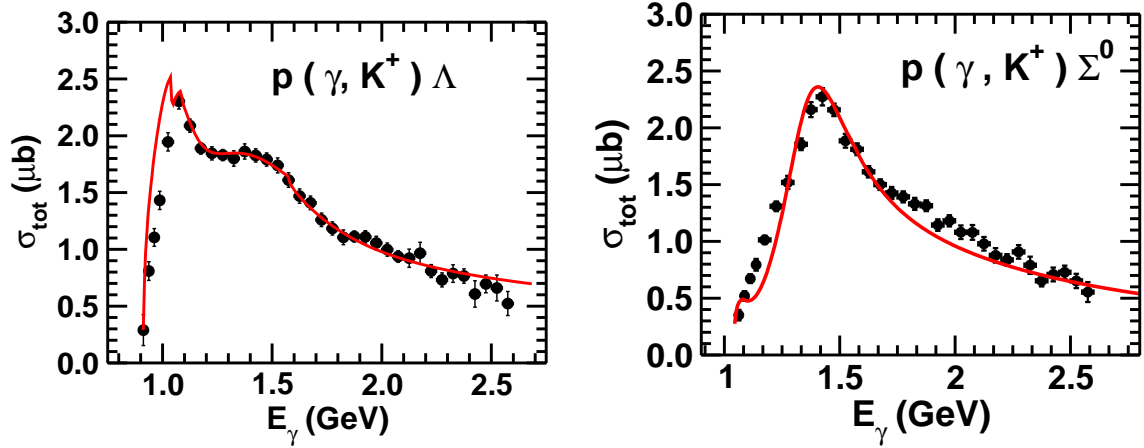

\begin{tabular}{cc}
\includegraphics[scale=0.60]{Fig2a.eps} & \hspace{0.30cm}
\includegraphics[scale=0.60]{Fig2b.eps}
\end{tabular}
\caption{[color online]
Comparison of the calculated total cross sections for the 
$p(\gamma,K^+)\Lambda$ and $p(\gamma,K^+)\Sigma^0$ reactions 
with the corresponding experimental data taken from
Ref.~\protect\cite{gla04}.}
\label{Fig2}
\end{figure*}

In Fig.~2, we compare the results of our calculations for the total
cross sections with the corresponding data of SAPHIR collaboration
\cite{gla04} for $p(\gamma,K^+)\Lambda$ and $p(\gamma, K^+)\Sigma^0$ 
reactions. Photon energies ($E_\gamma$) range from threshold to about 
2.6 GeV. The experimental cross sections for the $p(\gamma,K^+)\Lambda$ 
reaction show a steep rise as $E_\gamma$ increases from threshold 
to about 1.1~GeV. The latter corresponds to a $\gamma p$ channel total 
invariant mass ($W$) of $\approx$ 1.7 GeV which coincides with the 
masses of $S_{11}(1650)$, $P_{11}(1710)$ and $P_{13}(1720)$ resonances. 
The decrease in the cross sections just before the threshold for the 
$K^+\Sigma^0$ channel indicates a cusp due to the opening of this 
channel which has already been indicated in Refs.~\cite{kai97,uso05}. 
The data in this region are well reproduced by our calculations. 
There is also a second peak in the data at $E_\gamma \approx$ 1.5 
GeV ( $W$ $\approx$ 1.9 GeV). Our calculations describe 
the data well also in this region.
 
In contrast to studies reported in Ref.~\cite{mar06,mar01b,jan01,sar05},
we do not require an additional baryonic resonance $D_{13}$(1895) in 
order to explain the data in the second  peak region. Looking at the 
cross sections for the $p(\gamma, K^+)\Sigma^0$ reaction (right panel) 
one notices that the second maximum in the $p(\gamma,K^+)\Lambda$ data 
is centered around the same value of $W$ where the only peak is 
observed in the $K^+\Sigma^0$ channel. Hence, peaks in the 
$p(\gamma,K^+)\Lambda$ total cross section are more likely to be 
the consequences of unitarity and multi-channel dynamics. This is 
further supported by the  coupled channel analysis of this reaction 
as reported in Ref.~\cite{shk05} where the second peak in the 
$K^+\Lambda$ total cross section data is explained as resulting from 
the interference between background and resonance contributions and 
not due to the presence of a $D_{13}$(1895) resonant state. The same 
conclusion was arrived also in the previous coupled-channels studies
\cite{pen02}.

On the other hand, the total cross sections of the 
$p(\gamma, K^+)\Sigma^0$ reaction rise smoothly from threshold to its 
peak at $E_\gamma \approx$ 1.45 GeV ($W \approx$ 1.9 GeV). The cross 
sections drop smoothly for $E_\gamma$ larger than this value. Our 
calculations are able to reproduce the data well in the entire region of 
photon energies with the exception of some far lower and some higher 
photon energies where the data are somewhat underestimated.
\begin{figure*}
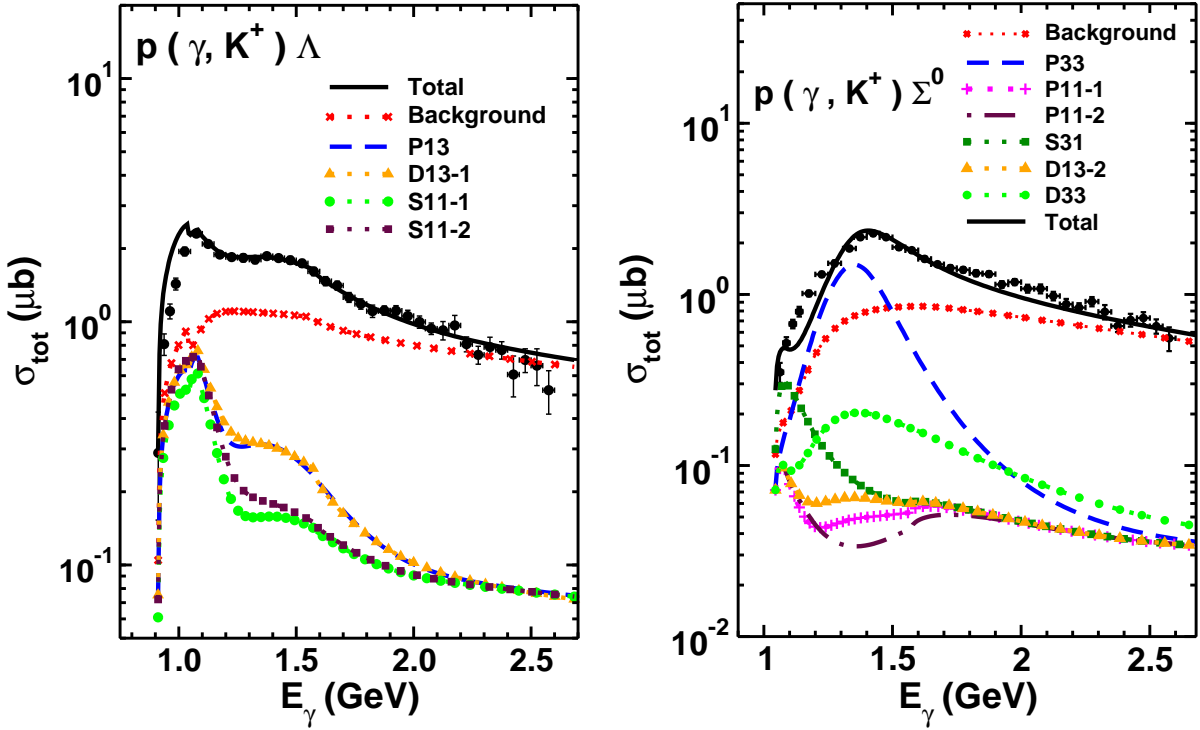

\begin{tabular}{cc}
\includegraphics[scale=0.60]{Fig3a.eps} & \hspace{0.20cm}
\includegraphics[scale=0.60]{Fig3b.eps}
\end{tabular}
\caption{[color online]
Partial wave decomposition of the calculated total cross sections for
the same reactions as those shown in Fig.~2.
Contributions of different resonances are shown by various curves
as indicated in the figure.  Also shown are the background contributions
which consist of Born  and $u$- and $t$-channel terms. Experimental
data are taken from Ref.~\protect\cite{gla04}.}
\label{Fig3}
\end{figure*}

In Fig.~3, we show the contribution of various resonances and  
background terms to the total cross sections of $p(\gamma,K^+)\Lambda$
and $p(\gamma, K^+)\Sigma^0$ reactions as a function of $E_\gamma$. 
It is clear from this figure that while the background contributions 
dominate the $p(\gamma,K^+)\Lambda$ cross sections in the entire range 
of photon energies, they do so only for $E_\gamma >$ 1.5 GeV  in 
case of the $p(\gamma, K^+)\Sigma^0$ reaction. It is interesting 
to note that contributions of $P_{13}(1720)$, $S_{11}(1535)$, 
$S_{11}(1650)$, $P_{13}(1720)$ and $D_{13}(1520)$ (depicted as 
$P13$, $S11$-1, $S11$-2, $D13$-1, respectively in the left panel of
Fig.~3) resonances peak at about the same value of $E_\gamma$ 
($\approx$ 1.1 GeV) in the $p(\gamma,K^+)\Lambda$ total cross section. 
In the region around this energy, the resonance contributions are 
comparable to those of the background terms and they combine together
to constitute the structure of the  first peak in the data. 
Furthermore, no one resonance individually dominates in this region 
which is in  contrast to the results of Ref.~\cite{shk05}. The 
contributions of the $P_{11}$(1710) resonance are very weak and 
are not included in our study. This is in agreement with the results
Ref.~\cite{shk05}. We have also not included a third $S_{11}$ resonance 
with mass and width around 1.780 GeV and 0.28 GeV, respectively which 
was considered in descriptions of $p(\gamma,\eta)p$ and 
$p(\gamma,K^+)\Lambda$ reactions in Refs.~\cite{he08,jul05}.  
\begin{figure*}
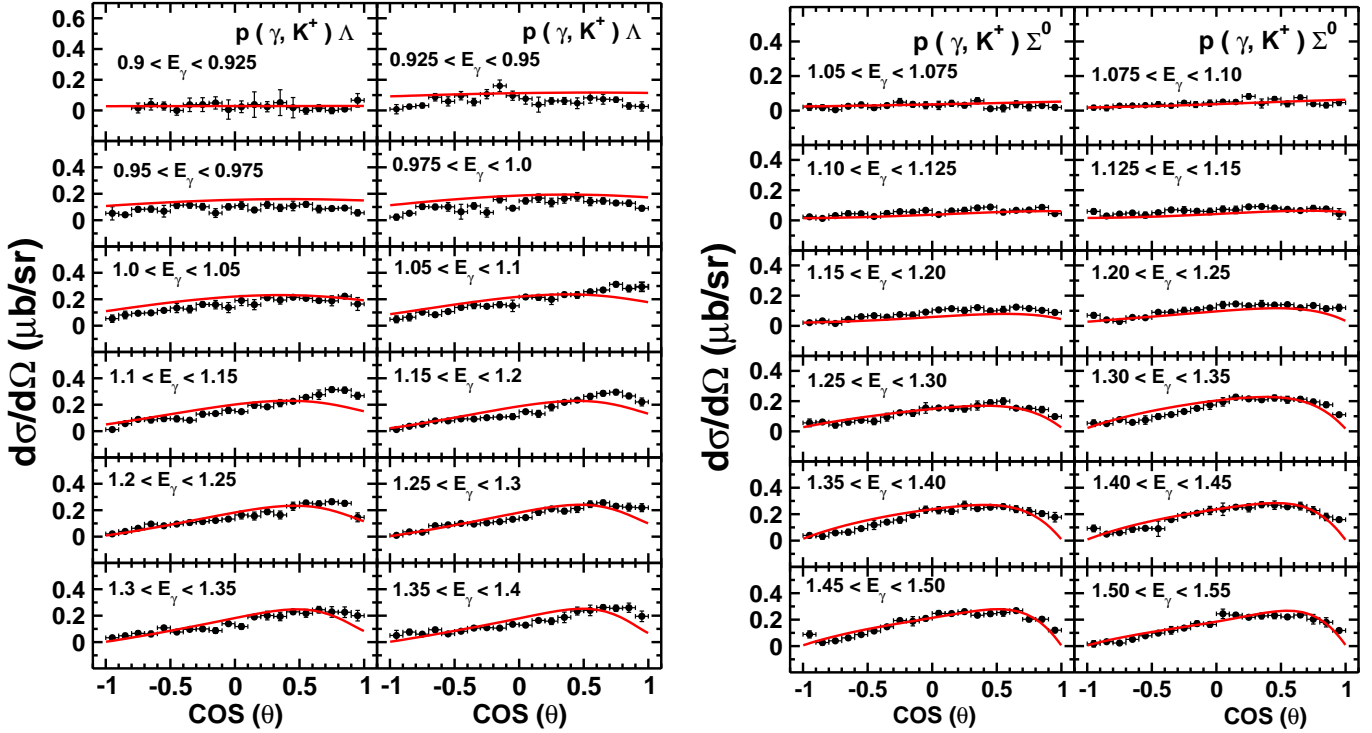

\begin{tabular}{cc}
\hspace{-1.2cm}\includegraphics[scale=0.50]{Fig4a.eps} & 
\hspace{0.20cm}
\includegraphics[scale=0.50]{Fig4b.eps}
\end{tabular}
\caption{[color online]
Comparison of the calculated and experimental differential cross 
sections for the  $p(\gamma,K^+)\Lambda$ and $p(\gamma,K^+)\Sigma^0$ 
reactions as a function of the cosine of the $K^+$ c.m.\ angle for 
photon energies 0.94 GeV $<$ $E_\gamma$ $<$ 1.4 GeV and and 1.05 GeV 
$<$ $E_\gamma$ $<$ 1.55 GeV, respectively. The energy bin 
is indicated in each graph in GeV. The experimental data are taken 
from Ref.~\protect\cite{gla04}.
}
\label{Fig4}
\end{figure*}

The peak region in the $p(\gamma, K^+)\Sigma^0$ total cross section, 
on the other hand, is dominated by the contributions from the 
spin-$\frac{3}{2}$, isospin-$\frac{3}{2}$ $P_{33}$(1600) resonance.
Apart from the background and to a lesser extent $D_{33}$(1700) 
terms other resonances are almost unimportant in this region.
Furthermore,  magnitudes of $D_{13}$(1700), $P_{11}$(1440) and  
$P_{11}$(1710) resonances (depicted as $D13$-2, $P11$-1 and $P11$-2, 
respectively in the right panel of Fig.~3) are comparatively small 
in the entire range of photon energies. However, for $E_\gamma$ 
very close to threshold the $S_{31}$(1620) resonance is most 
important. 

For both the reactions shown in Fig.~3, we note that the total cross 
sections beyond 2~GeV are almost solely governed by the
contributions of the background terms. In this region all resonance 
contributions are small and comparable to each other.
\begin{figure*}
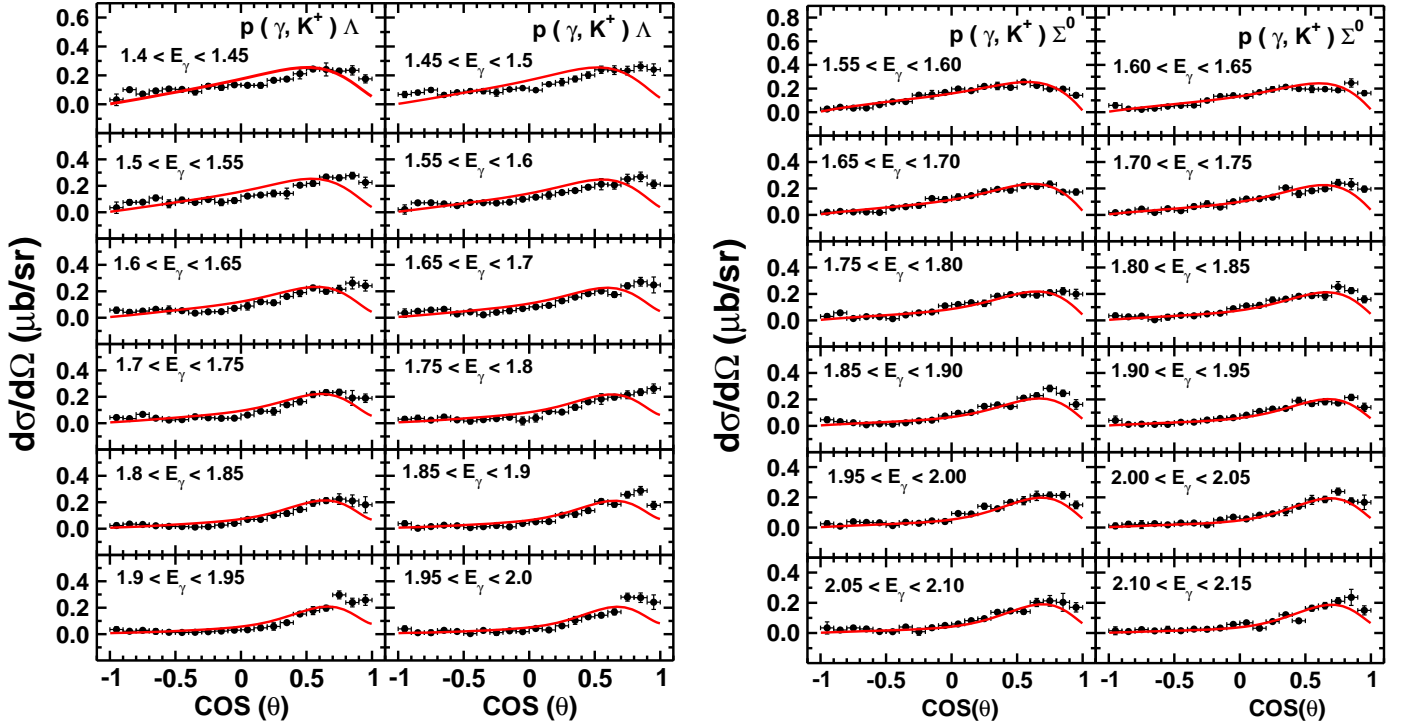

\begin{tabular}{cc}
\hspace{-1.2cm}\includegraphics[scale=0.50]{Fig5a.eps} & \hspace{0.20cm}
\includegraphics[scale=0.50]{Fig5b.eps}
\end{tabular}
\caption{[color online]
Same as that shown in Fig.~4 but for photon energies 
1.4 GeV $<$ $E_\gamma$ $<$ 2.0 GeV and 1.55 GeV 
$<$ $E_\gamma$ $<$ 2.15 GeV, respectively. The energy bin 
is indicated in each graph in GeV. The experimental data are 
taken from Ref.~\protect\cite{gla04}.
}
\label{Fig5}
\end{figure*}

Differential cross sections (DCS) provide more valuable information 
about the reaction mechanism. They reflect the quantum number of the
excited state (baryonic resonance) when the cross section is dominated 
by it. DCS include terms that weigh the interference terms of various 
components of the amplitude with the outgoing $K^+$ angles. Therefore, 
the structure of interference terms could highlight the contributions 
of different resonances in different angular regions. For   
$p(\gamma,K^+)\Lambda$ and $p(\gamma, K^+)\Sigma^0$ reactions DCS data 
of ELSA-SAPHIR group exist for 36 and 35 photon energy bins, respectively
in the range of respective thresholds to about 2.60~GeV covering a wide 
range of $K^+$ center of mass (c.m.) angles ~\cite{gla04}. In the left 
panels of Figs.~4-6 we show comparisons of our calculations for  
DCS with the corresponding SAPHIR data for the $p(\gamma,K^+)\Lambda$ 
reaction for energy bins in the range of 0.9~GeV -- 1.4~GeV, 
1.4~GeV -- 2.0~GeV, and 2.0~GeV -- 2.6 GeV, respectively, while in the 
right panels the same are shown for the $p(\gamma, K^+)\Sigma^0$ reaction 
for photon energy bins in the range of 1.05~GeV -- 1.55~GeV, 
1.55~GeV -- 2.15~GeV, and 2.15~GeV -- 2.55~GeV, respectively. 
\begin{figure*}
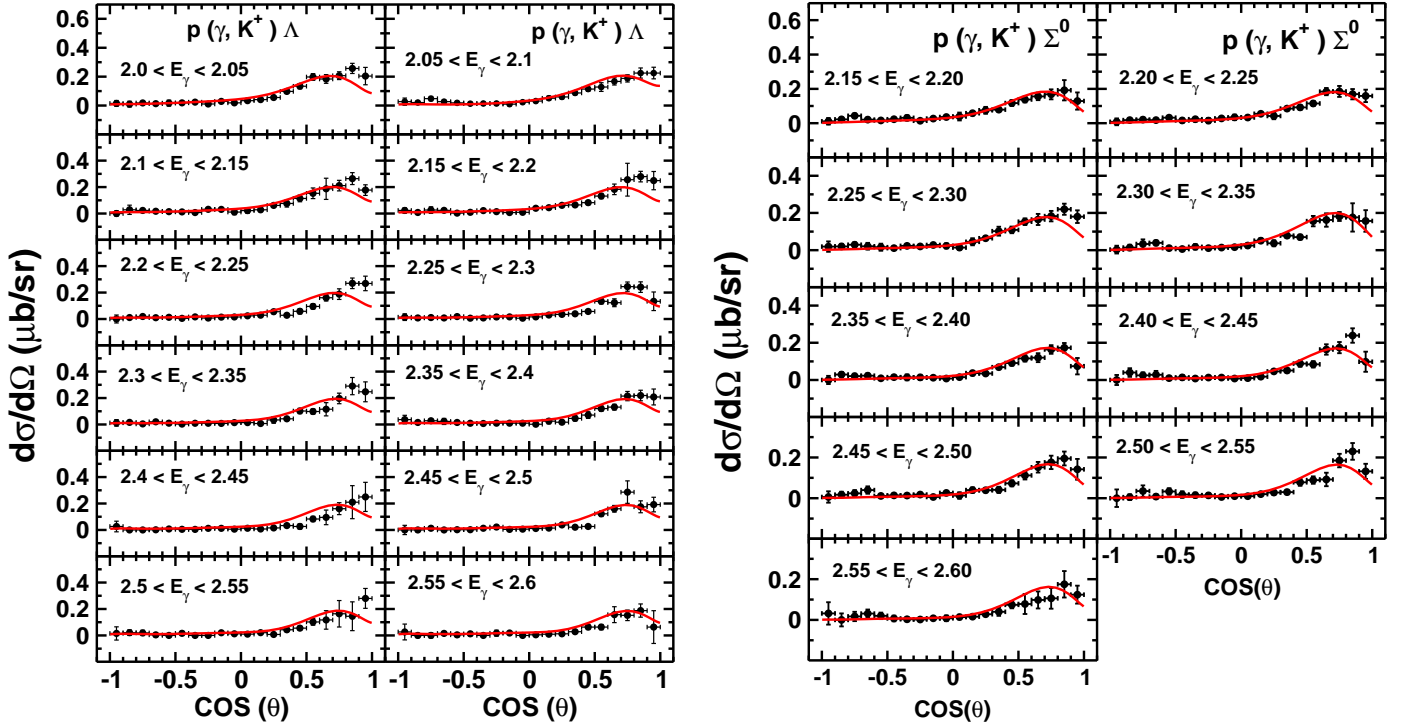

\begin{tabular}{cc}
\hspace{-1.2cm}\includegraphics[scale=0.50]{Fig6a.eps} & 
\hspace{0.20cm}
\includegraphics[scale=0.50]{Fig6b.eps}
\end{tabular}
\caption{[color online]
Same as that shown in Fig.~4 but for photon energies 
2.0 GeV $<$ $E_\gamma$ $<$ 2.6 GeV and 2.15 GeV 
$<$ $E_\gamma$ $<$ 2.55 GeV, respectively. The energy 
bin is indicated in each graph in GeV. The experimental data are 
taken from Ref.~\protect\cite{gla04}.
}
\label{Fig6}
\end{figure*}

It is seen that the differential cross sections are flat as a 
function of kaon angle near the respective thresholds which signifies the 
dominance of $S$-wave resonances near these energies. As $E_\gamma$ rises 
further the DCS develop a significant forward peaking which is 
consistent with the domination of the background terms or the interference 
between background and $s$-channel resonance contributions. At still higher 
energies, the data show a tendency of a slow rise at the extreme forward 
angles. 

It is clear that our model describes general trends of the data well 
in the complete photon energy regime of the SAPHIR measurement. However,
a few specific details of the data are missed for some energy bins. In 
the regions of 1.0--1.20~GeV and 1.45--1.8~GeV, the angular distributions 
at the extreme forward angles are not properly reproduced for the 
$p(\gamma,K^+)\Lambda$ reaction. However, those of the 
$p(\gamma, K^+)\Sigma^0$ reaction are well reproduced in these regions.
At photon energies $>$ 2.0~GeV, the $p(\gamma,K^+)\Lambda$ cross 
sections show a tendency of peaking at extreme forward angles while the 
$p(\gamma, K^+)\Sigma^0$ data do not seem to do so. Within statistical 
errors our calculations are consistent with this trend of the data, 
although for the $K^+\Lambda$ channel the agreement is of a lesser 
quality as compared to that for the $K^+\Sigma^0$ one. In view of 
the fact that our background terms include both $K$ and $K^*$ exchange 
diagrams with the same couplings for both the reactions, we have obtained
a reasonably good agreement with the data.   
\begin{figure*}
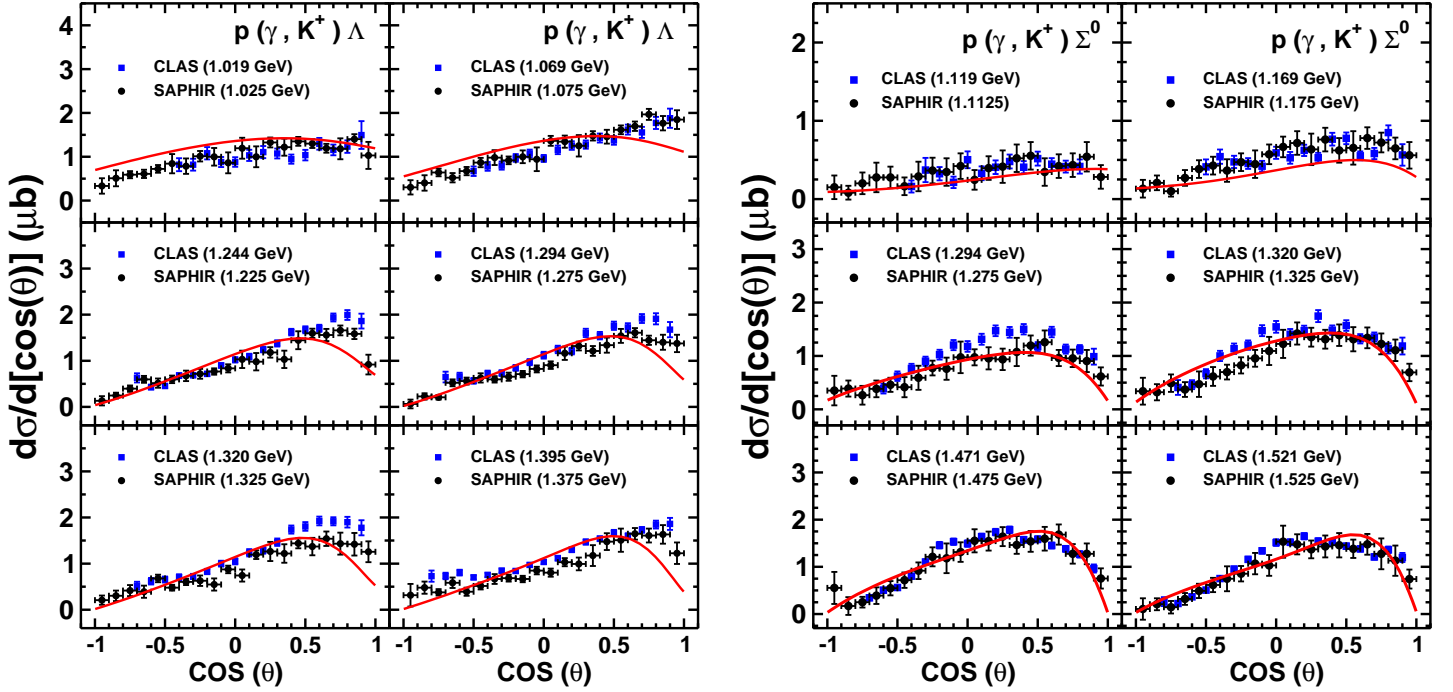

\begin{tabular}{cc}
\hspace{-1.2cm}\includegraphics[scale=0.50]{Fig7a.eps} & \hspace{0.24cm}
\includegraphics[scale=0.50]{Fig7b.eps}
\end{tabular}
\vskip -0.1in
\caption{[color online]
Comparison of the calculated and experimental differential cross 
sections from the CLAS collaboration~\protect\cite{bra06} (shown by 
solid blue squares) for the  $p(\gamma,K^+)\Lambda$ and 
$p(\gamma,K^+)\Sigma^0$ reactions as a function of the cosine of the 
$K^+$ c.m.\ angle for selected bins of photon energy 
$<$ 1.5. The data of the SAPHIR collaboration (taken from 
Ref.~\protect\cite{gla04}) at nearby 
energies are also shown for comparison (by solid black circles). 
}
\label{Fig7}
\end{figure*}

There are some discrepancies between the data of ELSA-SAPHIR and
CLAS collaborations. The CLAS group has reported consistently 
larger cross sections at most kaon angles for $E_\gamma >$ 1.19
GeV. For certain forward angles and photon energy bins one notices 
a large difference in the SAPHIR and CLAS data for DCS. Therefore,
a simultaneous description of the data of two collaborations has 
often been problematic for the theoretical models~\cite{mar06,jul06}.
Our model is no exception to this. We show in Figs.~7 and 8, 
comparisons between our calculations and the CLAS data for 
differential cross sections for $p(\gamma,K^+)\Lambda$ and 
$p(\gamma, K^+)\Sigma^0$ reactions for 12 chosen photon energy
bins. In these figures the SAPHIR data points are also given for 
the comparison purpose. Since, the CLAS data are given in energy and 
angular bins different from those of the SAPHIR one, we have chosen 
those bins which are nearly equal in two cases. 
\begin{figure*}
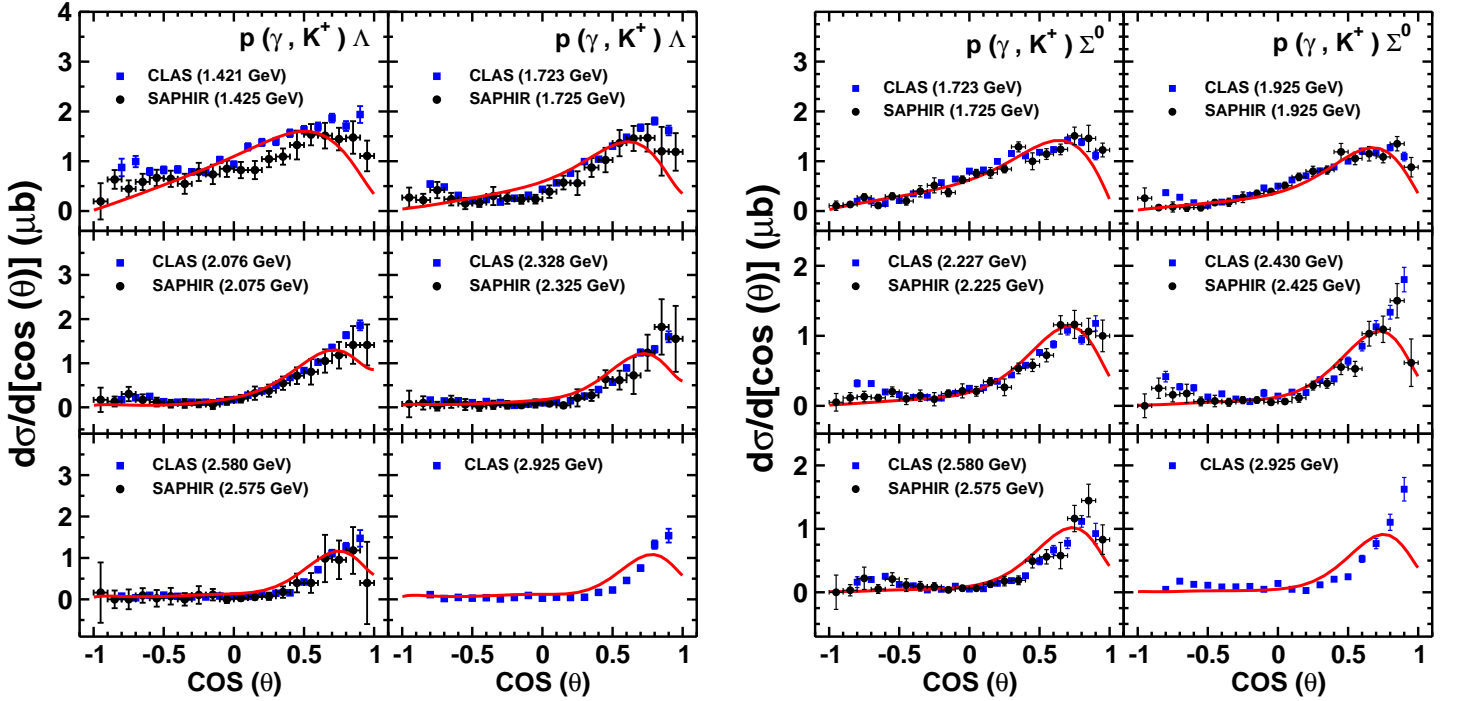

\begin{tabular}{cc}
\hspace{-1.2cm}\includegraphics[scale=0.50]{Fig8a.eps} & \hspace{0.32cm}
\includegraphics[scale=0.50]{Fig8b.eps}
\end{tabular}
\caption{[color online]
Same as that shown in Fig.~7 but for selected bins of photon energy 
$>$ 1.5 GeV.  
}
\label{Fig8}
\end{figure*}

We note that CLAS and SAPHIR data are identical for  $E_\gamma <$
1.1~GeV for both reactions. However, at larger energies considerable 
difference are seen between the two data sets particularly at forward 
angles. The CLAS DCS data for the $K^+\Lambda$ channel are more 
forward peaked for $E_\gamma$ $\approx$ 1.2 - 1.4~GeV as compared to 
the SAPHIR data as well as our calculated cross sections. For the 
$K^+\Sigma^0$ case, the CLAS DCS are consistently larger than the SAPHIR 
ones for all angles except for the backward ones for $E_\gamma$ between 
1.25 to 1.5 GeV. Our calculations are unable to reproduce this feature. 
 
For $E_\gamma$ between 1.4 -1.7 the $K^+\Lambda$ channel CLAS data
have the tendency of backward peaking as well along with the 
stronger forward peaking which is again in contrast to the SAPHIR data
as well as our calculations. The backward peaking of the CLAS data
disappears for photon energies between 2.0 - 2.6 GeV, however, stronger
forward peaking still remains and our calculations are unable to reproduce 
it fully. For the $p(\gamma,K^+)\Sigma^0$ reaction the CLAS data show
some backward peaking at photon energies between 2.2 - 2.4 GeV. 
At other energies the differences between the data of the 
two groups are less noticeable for this channel. Our calculations are 
in better agreement with the CLAS data for these cases.
 
\begin{figure*}
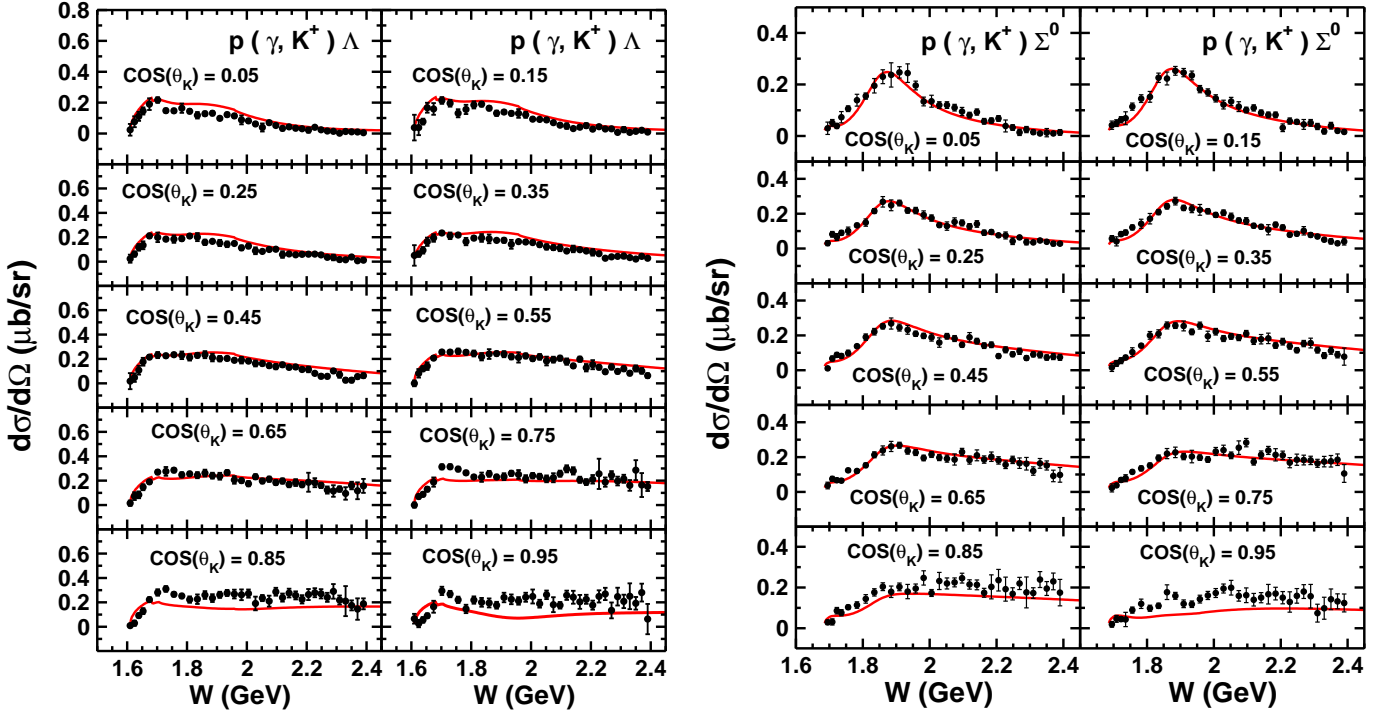

\begin{tabular}{cc}
\hspace{-1.0cm}\includegraphics[scale=0.50]{Fig9a.eps} & \hspace{0.20cm}
\includegraphics[scale=0.50]{Fig9b.eps}
\end{tabular}
\caption{[color online]
Comparison of the calculated and experimental differential cross 
sections for the  $p(\gamma,K^+)\Lambda$ and $p(\gamma,K^+)\Sigma^0$ 
reactions as a function of the $\gamma p$ invariant mass ($W$) for various 
positive cosines of the $K^+$ c.m.\ angle. The experimental data are 
taken from Ref.~\protect\cite{gla04}.
}
\label{Fig9}
\end{figure*}
Resonance structure in the $s$-channel should appear more clearly in 
the $W$ dependence of the differential cross sections at various 
$K^+$ angles.  In Fig.~9 (Fig.~10) we show the DCS for 
$p(\gamma,K^+)\Lambda$ and $p(\gamma,K^+)\Sigma^0$ reactions as a 
function of $W$ for positive (negative) values of COS$(\theta_{K^+})$. 
The experimental data are from the SAPHIR collaboration. From these
figures we note that overall shapes of the $W$ distributions are 
reproduced well by our calculations for all the angles for both the 
reactions. 

Nevertheless, we also notice that for the case of the $K^+\Lambda$ 
production our calculations underestimate the data somewhat for $W$ 
between 1.7-2.0 GeV at forward angles [for COS$(\theta_{K^+})$ = 0.75, 
0.85, and 0.95]. As bulk of the forward peaking is due to $t$-channel
exchanges which are well understood, this leads to the suggestion that a
contribution to the background is still missing. It this context 
supplementing the high energy parts of the $t-$channel exchanges by 
Regge-trajectory exchange~\cite{gui03,cor06,cor07a,cor07b,van09} may be 
an interesting option. Work is in progress to include the Regge-trajectory 
exchange in our model.
 
\begin{figure*}
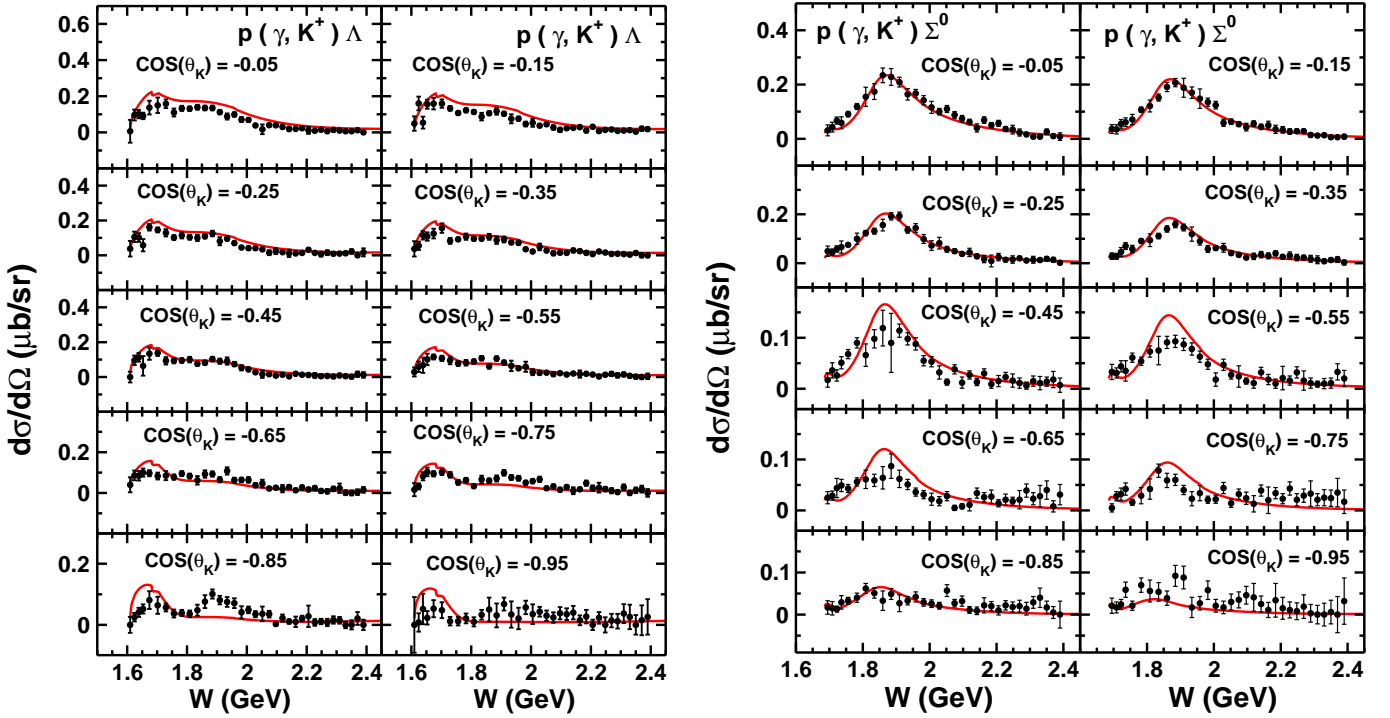

\begin{tabular}{cc}
\hspace{-1.0cm}\includegraphics[scale=0.50]{Fig10a.eps} 
& \hspace{0.20cm}
\includegraphics[scale=0.50]{Fig10b.eps}
\end{tabular}
\vskip -0.1in
\caption{[color online]
Same as that shown in Fig.~9 but for various negative cosines 
of the $K^+$ c.m.\ angle. The experimental data are taken from 
Ref.~\protect\cite{gla04}.
}
\label{Fig10}
\end{figure*}

For the backward angles (see Fig.~9), the data on the 
$p(\gamma,K^+)\Lambda$ reaction show two peaks at $W$ around 1.7 GeV 
and 1.9 GeV. Our calculations are able to reproduce the structure of 
both the peaks for all the angles except for the most backward one 
[COS$(\theta_{K^+})$ = 0.85] where the peak at $W \approx$ 1.9 GeV 
is underestimated.  We would like to stress that unlike 
Ref.~\cite{jul06}, we find no need of including an additional 
$D_{13}$ resonance (with mass $\approx$ 1.9 GeV and width 0.316 GeV) 
to reproduce the data around $90^0$.  The underestimation of the peak 
for $W$ around 1.9 GeV for COS$(\theta_{K^+})$ = 0.85 might be seen as 
an indication of the need to include such a resonance. However, with 
the inclusion of this resonance the cross section increases for other 
backward angles also around this value of $W$~\cite{mar01b} and the 
signs of the beam asymmetries come out to be opposite to what has 
been observed experimentally for this $W$~\cite{zeg03}.

On the other hand, for the $K^+\Sigma^0$ channel, the $W$ dependence 
of the DCR (see Figs.~9 and 10) is reproduced very well by our 
calculations for all the angles [corresponding to both negative and 
positive values of COS$(\theta_{K^+})$] with the exception of one very 
forward $\theta_{K^+}$. 
 
Polarization observables provide more sensitive tests of reaction 
models.  The reason for this lies in the fact that these observables 
are generally very sensitive to the imaginary parts of the amplitudes 
which are governed by coupling to other channels via the optical 
theorem. Data on hyperon recoil polarization ($P_Y$, with $Y$ being 
$\Lambda$ or $\Sigma^0$) have been reported by both SAPHIR and CLAS 
collaborations~\cite{gla04,mcn04}. $P_Y$ is related to interferences 
of the imaginary parts of the resonant amplitudes with real parts of 
other amplitudes including those of the back ground terms. In Fig.~11 
we compare the results of our calculations $P_Y$ with the corresponding 
data of the SAPHIR collaboration for $p(\gamma,K^+)\Lambda$ and 
$p(\gamma,K^+)\Sigma^0$ reactions. One notices that experimental 
$P_\Lambda$ tend to be positive at backward angles, nearly zero at 
angles around zero and negative at forward angles. In contrast, 
$P_{\Sigma^0}$ data show nearly opposite trend. The CLAS 
$P_Y$ data are used in the analyses of Refs.~\cite{jul06,mar01b}.
\begin{figure*}
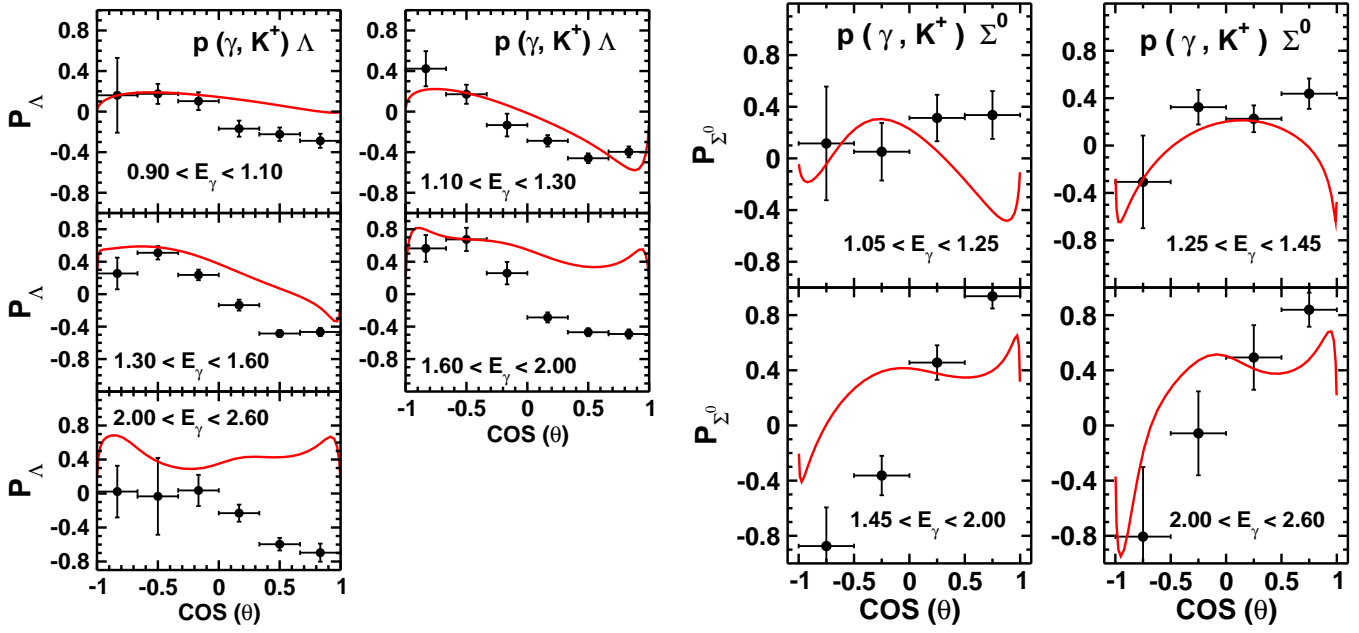

\begin{tabular}{cc}
\hspace{-1.2cm}\includegraphics[scale=0.50]{Fig11a.eps} & \hspace{0.20cm}
\includegraphics[scale=0.50]{Fig11b.eps}
\end{tabular}
\caption{[color online]
Comparison of our calculations with the experimental data (taken
from Ref.~\protect\cite{gla04}) for hyperon polarization for 
$p(\gamma,K^+)\Lambda$ and $p(\gamma,K^+)\Sigma^0$ reactions as a 
function of the cosine of $K^+$ c.m.\ angle for the photon energy 
bins (in GeV) as indicated in the figures.  
}
\end{figure*}
 
Our calculations reproduce approximately the trends seen in the 
data. The opposite signs of the observed $P_Y$ of two channels 
are nearly reproduced. The agreement with data is relatively better 
for $E_\gamma$ below 1.6 GeV. The large positive back angle 
polarizations  seen in the $P_\Lambda$ data are reproduced for all the 
photon energy bins. Similarly the large negative experimental 
$P_{\Sigma^0}$ at these angles are also nearly reproduced. At 
forward angles our model is relatively less successful in reproducing
the $P_\Lambda$ data at $E_\gamma$ above 1.6 GeV. This again indicates
perhaps some inadequacy of the background terms within our model.  

Beam asymmetry ($\Sigma_B$) is the measure of the azimuthal anisotropy 
of a reaction yield relative to the linear polarization of the incoming 
photon. In Fig.~12 we compare the results of our calculations for this 
observable for both $p(\gamma,K^+)\Lambda$ and $p(\gamma,K^+)\Sigma^0$
reactions with the corresponding data taken from Ref.~\cite{zeg03,sum06}  
which are available for nine energy bins for photon energies between 1.5
GeV to 2.3 GeV. Available experimental $\Sigma_B$ are positive for both
the reactions. Our calculations reproduce the data with varying degrees
of success. For the $p(\gamma,K^+)\Lambda$ reaction the agreement 
with the data is relatively better for $E_\gamma <$ 2.0 GeV. For 
larger $E_\gamma$ the data are underestimated by our model. On the 
other hand, for the $p(\gamma,K^+)\Sigma^0$ case the data are described 
better for $E_\gamma >$ 1.9 GeV while at lower photon energies they are
overestimate. In calculations reported in Ref.~\cite{shk05} comparison 
with the data is shown for the $K^+\Lambda$ channel for one photon 
energy ($E_\gamma =$ 1.946 GeV) only. In the coupled-channels analysis 
of Ref.~\cite{jul06}, the agreement with the $\Sigma_B$ data for the 
$p(\gamma,K^+)\Lambda$ reactions is of the same quality as that achieved 
by us -- there too the data are better reproduced for $E_\gamma <$ 2.0 
GeV while they are underpredicted at energies larger than this. However, 
we emphasize that such data for both $K^+\Lambda$ and $K^+\Sigma^0$ 
channels have not been simultaneously analyzed with one parameter set 
in any other coupled-channels model.    
\begin{figure*}
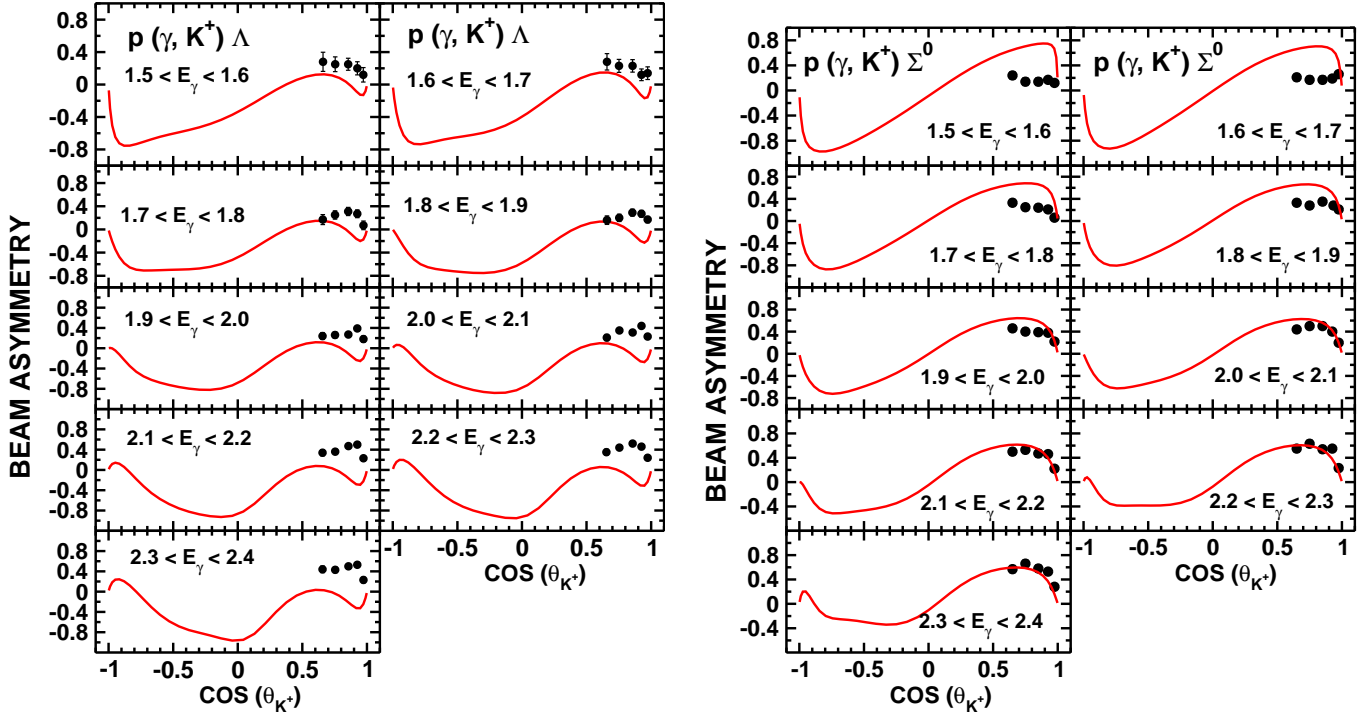

\begin{tabular}{cc}
\hspace{-1.2cm}\includegraphics[scale=0.50]{Fig12a.eps} & \hspace{0.20cm}
\includegraphics[scale=0.50]{Fig12b.eps}
\end{tabular}
\vskip -0.1in
\caption{[color online]
Beam asymmetry for the $p(\gamma,K^+)\Lambda$ and $p(\gamma,K^+)\Sigma^0$
reactions as a function of the cosine of $K^+$ c.m.\ angle for 15 photon 
energies. The experimental data are taken from Ref.~\protect\cite{zeg03}.
}
\end{figure*}

\section{Summary and Conclusions}

In this paper we investigated the photoproduction reactions 
$p(\gamma,K^+)\Lambda$ and $p(\gamma,K^+)\Sigma^0$  
within a coupled-channels effective-Lagrangian approach which 
is based on the \K-matrix method. Unitarity effects are correctly taken
into account, since all important final channels (consisting of two-body
systems $\pi N$, $\eta N$, $\phi N$, $\rho N$, $\gamma N$, $K \Lambda$, 
and $K \Sigma$) are included in the \K-matrix kernel. We build this 
kernel by using effective Lagrangians for the Born, $u$-channel, 
$t$-channel, and spin-$\frac{1}{2}$ and spin-$\frac{3}{2}$ resonance 
contributions. Thus, the background contributions are generated 
consistently and crossing symmetry is obeyed. The advantage of a full 
coupled-channel calculation is that it allows for the simultaneous
calculation of observables for a large multitude of reactions with 
considerably fewer parameters than would be necessary if each reaction 
channel is fitted separately. More significantly, the implementation 
of unitarity ensures that the imaginary parts of the amplitudes are 
compatible with the cross sections for other channels.

Our model provides reasonable description of the experimental data 
of the SAPHIR group on total and differential cross sections as well 
as on hyperon polarizations for photon energies ranging from threshold 
to up to $\approx$ 3~GeV for both $p(\gamma,K^+)\Lambda$ and 
$p(\gamma,K^+)\Sigma^0$ reactions. The beam asymmetry data of the 
SPring8/LEPS group for the two reactions are also reasonably reproduced.
 
An important point about our study is that the same parameter set was 
used in calculating observables for both the reactions. 

We made a detailed investigation of the contribution of various partial
waves and showed that peaks seen in the total $p(\gamma,K^+)\Lambda$ 
cross sections for photon energies around 1.5 GeV (invariant mass of 
1.9 GeV) are largely due to coupled-channels effects rather than 
the contribution of a $D$-wave resonance with mass around 1.9 GeV. 
A major part of cross sections of this reaction is generated via the 
background  terms and resonances give prominent contributions only in 
the first peak region (for $E_\gamma$ around 1.1 GeV). On the other hand,
the total $p(\gamma,K^+)\Sigma^0$ cross sections are dominated by the 
$P_{33}$ (1600) resonance in its peak region ($E_\gamma$ around 1.5 GeV).
 
The agreement between our calculations and the differential cross 
section data of the CLAS collaboration is not of the same quality
as that seen in case of the SAPHIR data, particularly at extreme 
forward angles. At some backward angles the CLAS data show pronounced 
peaks for some photon energies but no peak is seen in the corresponding 
SAPHIR data or in our theoretical cross sections. Whether or not 
these peaks are an indication for an additional $P$ or $D$ wave resonance 
is a matter of debate and additional investigations. In this 
context, it is quite desirable to settle the issue of mutual 
inconsistency between the CLAS and SAPHIR data sets. 

Our work shows that it is indeed possible to fit meson photoproduction
data of many channels simultaneously with a single parameter set within
a coupled-channels model. Further improvement of our model is however,
necessary for a better description of the data at extreme forward
angles in case of higher photon energies.

\section{Acknowledgments}
This work has been supported by Sonderforschungsbereich/Transregio 16, 
Bonn--Giessen--Bochum of the German Research Foundation (DFG) and
Dutch Organization for Scientific Research (NWO). 

\appendix
\section{Effective Lagrangians}

We list here the effective Lagrangians for various vertices. $p$, $k$,
$p\prime$ and $-q$ represent four momenta of the initial nucleon, 
final meson, final nucleon and photon, respectively. We assume that 
meson momenta are directed into the vertex, so that energy momentum 
conservation reads as $p + k = p\prime - q $.

For the nucleon vertices the following couplings were used
\begin{equation}\eqlab{LagrangianN}
  \begin{aligned}
  \La_{NN\pi} &= i g_{NN\pi}{\bar{\Psi}}_N \YYv{\vecvarphi_\pi}\Psi_N \\
  \La_{NN\eta} &= i g_{NN\eta}{\bar{\Psi}}_N \YY{\varphi_\eta} \Psi_N \\
  \La_{NN\sigma} &= -g_{NN\sigma}{\bar{\Psi}}_N \varphi_\sigma \Psi_N \\
 \La_{NN\rho} &= -g_{NN\rho}{\bar{\Psi}}_N \XXv{\vecvarphi_\rho\,} \Psi_N
\\
\La_{NN\omega} &= -g_{NN\omega}{\bar{\Psi}}_N \XX{\varphi_\omega}\Psi_N \\
\La_{NN\phi} &= -g_{NN\phi}{\bar{\Psi}}_N \XXp{\varphi_\phi}\Psi_N \\
\La_{NN\gamma} &= -e {\bar{\Psi}}_N \left( \frac{1+\tau_0}{2}
                    \gamma_\mu A^\mu +\frac{\kappa_\tau}{2m_N}
                    \sigma_{\mu\nu} \partial^\nu A^\mu \right)\Psi_N \\
\La_{NN\gamma\varphi}&= -e\frac{g_{NN\phi}}{2m_N}{\bar{\Psi}_N} \gamma_5
                         \gamma_\mu[\tauiso \times \vecvarphi]A^\mu \;.
\end{aligned}
\end{equation}
The parameter $\chi$ controls the admixture of pseudoscalar and 
pseudovector components in the corresponding Lagrangian. Its value is 
taken to be 0.5. This value was obtained in our previous study of 
photoproduction of associated strangeness~\cite{uso05} and has been 
held fixed in the study of all other reactions within our model. 
Nucleon spinors are depicted by $\Psi$ and meson fields by $\varphi$. 
The magnetic moments are represented by $\kappa$. 
$\La_{NN\gamma\varphi}$ generates the seagull or the contact term
diagrams. We have followed the notations of Ref.~\cite{bjo64}.

The Lagrangians for the meson vertices are
\begin{equation}\eqlab{LagrangianM}
  \begin{aligned}
   \La_{\rho\pi\pi} &= - g_{\rho\pi\pi}
         {\vecvarphi_\rho}_\mu \cdot (\vecvarphi_\pi \times \lrpart^\mu
          \vecvarphi_\pi) /2 \\
    \La_{\gamma\pi\pi} &= e \eps_{3ij} A_\mu
       (\varphi_{\pi_i} \lrpart^\mu \varphi_{\pi j}) \\
    \La_{\rho\gamma\pi} &= e \frac{g_{\rho\gamma\pi}}{m_\pi}\dop
     {\vecvarphi_\pi}{ \EPS{\partial}{A}{\partial}{\vecvarphi_\rho\,}} \\
    \La_{\omega\gamma\pi} &= e \frac{g_{\omega\gamma\pi}}{m_\pi}
                \varphi_{\pi^0}\EPS{\partial}{A}{\partial}{\omega} \\
\La_{\phi\gamma\pi} &= e \frac{g_{\phi\gamma\pi}}{m_\pi} 
\vecvarphi_{\pi^0}
                   \EPS{\partial}{A}{\partial}{\phi} \\
\La_{\phi\gamma\eta} &= e \frac{g_{\phi\gamma\eta}}{m_\pi} \varphi_\eta
        \EPS{\partial}{A}{\partial}{\varphi_\phi} \\
\La_{\rho\gamma\eta} &= e \frac{g_{\rho\gamma\eta}}{m_\pi} \varphi_\eta
       \EPS{\partial}{A}{\partial}{{\varphi_{\rho^0}}} \\
   \La_{\rho\gamma\sigma} &= e \frac{g_{\rho\gamma\sigma}}{m_\rho}
        (\partial^\mu \varphi_{\rho^\nu} \partial_\mu A_\nu
          - \partial^\mu \varphi_{\rho^\nu} \partial_\nu A_\mu ) \\
    \La_{\rho\rho\gamma} &= 2e \big(
    A^\mu (\partial_\mu {\varphi_\rho}_\nu) \tau_0 {\varphi_\rho}^\nu
    -(\partial^\nu A^\mu) {\varphi_\rho}_\nu \tau_0 {\varphi_\rho}_\mu \\
      &\quad +(\partial^\nu A^\mu) {\varphi_\rho}_\mu \tau_0
     {\varphi_\rho}_\nu \big)\\
  \La_{\phi \K \K} &= -ig_{\phi \K \K}{\bar \vecvarphi_\K} \lrpart^\mu
                      \vecvarphi_\K) \phi_\mu\\
  \La_{\eta\K^*\K}&= -i g_{\eta\K\K^*} \vecvarphi_\K \lrpart^\mu
                     \varphi_\eta {\bar {\vecvarphi_{\K^*}}}_\mu\\
  \La_{\pi\K^*\K}& = -i g_{\pi\K\K^*} {\bar \vecvarphi_\K} \lrpart^\mu
                  \vecvarphi_\pi \cdot \tauiso {\vecvarphi_{\K^*}}_\mu \\
   \La_{\rho\pi\eta}&= -i g_{\rho\pi\eta} (\varphi_\eta \lrpart^\mu
                       \vecvarphi_\pi) {\vecvarphi_\rho}_\mu\\
  \La_{\K^*\K^0\gamma}& = \frac{g_{\K^*\K\gamma}}{m_\pi}
                {\bar \vecvarphi_{\K^0}}
       \EPS {\partial}{A}{\partial}{\vecvarphi_{\K^*}}\\
    \La_{\K^*\K^\pm\gamma}& = \frac{g_{\K^*\K\gamma}}{m_\pi}
                  {\bar \vecvarphi_{\K^\pm}}
       \EPS {\partial}{A}{\partial}{\vecvarphi_{\K^*}} \;.
  \end{aligned}
\end{equation}
The coupling constants entering into Eqs. (A.1) and (A.2) together 
with baryon magnetic moments are listed in Table III. We have taken 
positive values for all primary coupling constants involving the 
nucleon. In particular, the sign of $g_{NK \Lambda}$ differs from 
its customary negative value~\cite{cot04}  However, we would like 
to stress that in a calculation like ours and also in many of those 
cited in Ref.~\cite{cot04} this sign is undetermined. Changing the 
sign of all the coupling constants involving a single $\Lambda$-field 
leaves the calculated observables invariant since it corresponds to a 
sign redefinition of this field. The magnitudes of the couplings are 
within the broad range specified in~\cite{cot04}.
\begin{table}
  \caption{\tbllab{parameters} Parameters summary table}
  \begin{ruledtabular}
  \begin{tabular}{C|C|C|C}
    g$_{NN\pi}$            & 13.47   & g$_{NN\eta}$           &  0.85  \\
    g$_{NN\sigma}$         & 10.0    & g$_{NN\rho}$           &  4.2   \\
    g$_{NN\omega}$         &  3.0    & g$_{NN\phi}$           & -0.0   \\
    g$_{\Sigma\sigma\rho}$ & 33.0    & g$_{\Sigma\Lambda\rho}$&-27.0   \\
    g$_{\rho\pi\pi}$       & 6.0     & g$_{\rho \pi \eta}$    &  0.0   \\
    g$_{\rho\pi^0\gamma}$ & -0.12    & g$_{\rho\pi^\pm\gamma}$& -0.10  \\
    g$_{\rho\eta\gamma}$  & -0.21    & g$_{\omega \eta\gamma}$& -0.12  \\
    g$_{\omega\pi\gamma}$  & 0.32    & g$_{\rho\sigma\gamma}$ & 12.0   \\
    g$_{N\Lambda \K}$      & 10.0    & g$_{N\Sigma K}$        & 14.5   \\
    g$_{N\Lambda \K^*}$    & -3.3    & g$_{N\Sigma K^*}$      &  0.0   \\
    g$_{\phi\K \K}$        & -4.5    & g$_{\rho \K \K}$       & -3.0   \\
    g$_{\pi \K \K^*}$      & -3.26   & g$_{\eta \K \K^*}$     & -3.2   \\
    g$_{\K^*\K^0\gamma}$   & 0.177   & g$_{\K^*\K^\pm\gamma}$ & -0.177 \\
    $\kappa_p$             & 1.79    & $\kappa_n$             & -1.91  \\
    $\kappa_\Lambda$       &-0.613   & $\kappa_{\Sigma^0}$    &  0.79  \\
    $\kappa_{\Sigma^+}$    & 1.45    & $\kappa_{\Sigma^-}$    & -0.16  \\
    $\kappa_{\Sigma^0 \rightarrow \Lambda \gamma}$ & -1.61    &        \\
  \end{tabular}
  \end{ruledtabular}
\end{table}

For the $S_{11}$, $S_{31}$,$P_{11}$ and $P_{3,1}$ resonances the hadronic
couplings are written as
\begin{eqnarray}
\La_{\varphi NR_{1/2}} & = & -g_{\varphi NR} {\bar{\Psi}}_R [\chi{i\Gamma}
        {\varphi}+(1-\chi)\frac{1}{M}\Gamma \gamma_\mu(\partial^\mu 
\varphi)]
                  \Psi_N + {\rm H.c.},
\end{eqnarray}
where $M \,=\,(m_R \,\pm\,m_N)$, with upper sign for even parity and
lower sign for odd parity resonance. 
The operator $\Gamma$ is $\gamma_5$ and unity for even and odd parity
resonances, respectively.  For isovector mesons, $\varphi$ in Eq.~(A.3)
needs to be replaced by $\tauiso \cdot \vecvarphi$ for 
isospin-$\frac{1}{2}$ resonances and by $\bf T \cdot \vecvarphi$ 
otherwise.

The corresponding electromagnetic couplings are
\begin{eqnarray}
\La_{\gamma NR_{1/2}} & = & -eg_1 {\bar{\Psi}_R} \frac{\Gamma}{4m_N}
                          \sigma_{\mu\nu}\Psi_N F^{\mu\nu} + H.c.,
\end{eqnarray}
where $\Psi_R$ is the resonance spinor and 
$F^{\mu\nu} = \partial^\mu A^\nu - \partial^\nu \A^\mu$. 
The operator $\Gamma$ is 1 for the positive
parity resonance and $-i\gamma_5$ for the negative parity one.

For spin-$\frac{3}{2}$ resonances, we have used the gauge-invariant effective
Lagrangians as discussed in Refs.~\cite{kon00,pas00,pas01,pas98,luk06}.
We write here the vertex functions used by us in computation involving 
these vertices. The resonance-nucleon-pion vertex function (e.g.) is 
given by 
\begin{eqnarray}
\Gamma_{R_{3/2} \to N\pi}^\alpha &=& {\frac{g_1}{m_\pi}}\,
   \Big[ \gamma^\alpha (q \cdot p) -{p\!\!\!/} q^\alpha \Big]
    [(1-\chi) + \chi {p\!\!\!/} /M_p],
\end{eqnarray}
and the corresponding electromagnetic vertices are
\begin{eqnarray}
\Gamma_{R_{3/2}\to N\gamma}^{\alpha \mu} &=&   \Bigg{\{}
  (g_2 + 2 g_1) \Big[ q^\alpha p^\mu -g^{\alpha\mu} p\cdot q \Big] +
     \nonumber \\
   && g_1 \Big[ g^{\alpha\mu}{p\!\!\!/}{q\!\!\!/} - q^\alpha {p\!\!\!/}
\gamma^\mu + \gamma^\alpha(\gamma^\mu p\cdot q - p^\mu {q\!\!\!/}) \Big]
      + \nonumber \\
   && g_3 \Big[(-q^2 g^{\alpha\mu} + q^\mu q^\alpha){p\!\!\!/} +
             (q^2 p^\mu - q^\mu p\cdot q) \gamma^\alpha \Big]
   \Bigg{\}} \nonumber \\
   &&\times \gamma_5 [(1-\chi) + \chi {p\!\!\!/} /M_p].
\end{eqnarray}
Here $p$ is the four-momentum of the resonance and $q$ is that of the 
meson. Index $\alpha$ belongs to the spin-$\frac{3}{2}$ spinor and 
$\mu$ to photon. Interesting property of these vertices is that the 
product, $p \cdot \Gamma$ = 0, where $\Gamma$ defines the vertices on 
the left hand side of Eqs.~(A.5) and (A.6).  As a consequence, the 
spin-$\frac{1}{2}$ part of the corresponding propagator becomes 
redundant as its every term is proportional to either 
$p_\mu$ or $p_\nu$. Thus only spin-$\frac{3}{2}$ part of this propagator
gives rise to non-vanishing matrix elements.


\begin{thebibliography}{99}

\bm{bra06}
R. Bradford {\it et al.}, Phys. Rev. C{\bf 73}, 035202 (2006), R. A. 
Schumacher, Private communication (2009).

\bm{mcn04}
J.W.C. McNabb {\it et al.}, Phys. Rev. C {\bf 69}, 042201 (2004).

\bm{gla04}
K.H. Glander {\it et al.}, Eur. Phys. J. A {\bf 19}, 251 (2004).

\bm{law05}
R. Lawall {\it et al.}, Eur. Phys. J. A {\bf 24}, 275 (2005).

\bm{zeg03}
R.G.T. Zengers {\it et al.}. Phys. Rev. Lett. {\bf 91}, 092001 (2003).

\bm{sum06}
M. Sumihama {\it et al.}, Phys. Rev. C {\bf 73}, 035214 (2006).

\bm{bra07}
R. Bradford {\it et al.}, Phys. Rev. C{\bf 75}, 035205 (2007)

\bm{cap00}
S. Capstick and W. Roberts, Prog. Part. Nucl. Phys. {\bf 45}, 241 
(2000) and references therein.

\bm{and09}
Andre Walker-Loud, Huey-Wen Lin, D.G. Richards, R.G. Edwards, 
M. Engelhardt, G.T. Flemming, Ph. Hagler, B. Musch, M.F. Lin, 
Harvey B. Meyer, John W. Negele, A.V. Pochinsky, Massimiliano Procura, 
Sergey Syritsyn, C.J. Morningstar, K. Orginos, D.B. Renner, 
W. Schroers, Phys.Rev.D {\bf 79}, 054502 (2009).

\bm{bas07}
S. Basak, R. G. Edwards, G. T. Fleming, K. J. Juge, A. Lichtl, 
C. Morningstar, D. G. Richards, I. Sato, S. J. Wallace, 
Phys.Rev.D {\bf 76}, 074504 (2007).

\bm{bur06}
T. Burch, C. Gattringer, L. Ya. Glozman, C. Hagen, D. Hierl, 
C. B. Lang, and A. Sch\"afer, Phys. Rev. D. {\bf 74}, 014504 (2006).

\bm{mat05}
N. Mathur, Y. Chen, S.J. Dong, T. Draper, I. Horváth, F.X. Lee, 
K.F. Liu, and J.B. Zhang, Phys. Lett. {\bf B605}, 137 (2005). 
  
\bm{lei05}
D. B. Leinweber, W. Melnitchouk, D. G. Richards, A. G. Williams, 
and L. M. Zanotti, in {\it Lattice Hadron Physics}, edited by A. 
Kalloniatis, D.  Leinweber, A. Williams (Springer, Berlin, 2005), 
P. 71; J. M. Zanotti, B. Lasscock, D. B. Leinweber and A. G. Williams,
Phys. Rev.  {\bf D71}, 034510 (2005); R.D. Young, D. B. Leinweber, 
and A. W. Thomas, Phys. Rev. {\bf D71}, 014001 (2005).

\bm{lor01}
U. L\"oring, K. Kretzschmar, B. C. Metsch, H. R. Petry, 
Eur. Phys.  J A {\bf 10}, 395 (201).

\bm{lut05}
M. F. M. Lutz and E. E. Kolomeitsev, Nucl. Phys. {\bf A755}, 29 (2005);
J. Hofmann and M. F. M. Lutz, Nucl. Phys. {\bf A763}, 90 (2005); J. Hofmann and
M. F. M. Lutz, Nucl. Phys. {\bf A776}, 17 (2006)

\bm{ben95}
M. Benmerrouche, N. C. Mukhopadhyay, J. F. Zhang, Phys. Rev. D {\bf 51},
3237 (1995); N. C. Mukhopadhyay and N. Mathur, Phys. Lett. {\bf B444},
7 (1998); R. M. Davidson, N. Mathur, N. C. Mukhopadhyay, Phys. Rev. C 
{\bf 62}, 058201 (2000).

\bm{arn00}
R. Workman, R. A. Arndt, I. I. Strakovsky, Phys. Rev. C {\bf 62}, 048201
(2000); R. A. Arndt, W. J. Briscoe, I. I. Strakovsky, R. L. Workman,
Phys. Rev. C {\bf 66}, 055213; Ya. I. Azimov, R. A. Arndt, 
I. I. Strakovsky, and R. L. Workman, Phys. Rev. C {\bf 68}, 045204 
(2003); R. A. Arndt, W. J. Briscoe, I. I. Strakovsky, and R. L. Workman,
Phys.Rev. C {\bf 72}, 058203 (2005).

\bm{tia99}
L. Tiator, D. Drechsel, G. Kn\"ochlein, and C. Bennhold, 
Phys. Rev. C {\bf 60}, 035210 (1999).

\bm{dre99}
D. Drechsel, O. Hanstein, S. S. Kamalov, L. Tiator, Nucl. Phys.
{\bf A 645}, 145 (1999)

\bm{shy07} R. Shyam, Phys. Rev. C {\bf 75}, 055201 (2007); 
R. Shyam, Phys. Rev. C {\bf 60}, 055213 (1999); 
R. Shyam, G. Penner and U. Mosel, Phys. Rev. C {\bf 63}, 022202(R) 
(2001); 
R. Shyam, Phys. Rev. C {\bf 73}, 035211 (2006).

\bm{jan01}
Stijn Janssen, Jan Ryckebusch, Dimitri Debruyne, and Tim Van Cauteren,
Phys. Rev. C {\bf 65}, 015201 (2001).

\bm{feu98}
T. Feuster and U. Mosel, Phys. Rev.C {\bf 58}, 457 (1998); 
T. Feuster and U Mosel, Phys. Rev.C {\bf 59},460 (1999).

\bm{kor98}
A. Yu. Korchin, O. Scholten, and R. G. E. Timmermans, 
Phys. Lett. {\bf B438}, 1 (1998).

\bm{pen02}
G. Penner and U. Mosel, Phys. Rev. C {\bf 66}, 055211 (2002); 
{\bf 66}, 055212 (2002); G. Penner, Ph.D. thesis (in English), 
Universit\"at Giessen, 2002, available at the URL 
http://theorie.physik.uni-giessen.de

\bm{uso05}
A. Usov and O. Scholten, Phys. Rev. C{\bf 72}, 025205 (2005).

\bm{shk05}
V. Shklyar, H. Lenske, and U. Mosel, Phys. Rev. C {\bf 72},
015210 (2005).

\bm{uso06}
A. Usov and O. Scholten, Phys. Rev. C{\bf 74}, 015205 (2006).

\bm{shy08}
R. Shyam and O. Scholten, Phys. Rev. C {\bf 78}, 065201 (2008).

\bm{dur08}
J. Durand, B. Julia-Diaz, T.-S. H. Lee, B. Saghai, T. Sato, Phys.Rev. C
{\bf 78}, 025204 (2008).
 
\bm{jul08}
B. Julia-Diaz, T.-S. H. Lee, A. Matsuyama, T. Sato, L. C. Smith, 
Phys. Rev. C {\bf 77}, 045205 (2008).

\bm{jul07}
B. Julia-Diaz, T.-S. H. Lee, A. Matsuyama, T. Sato, Phys. Rev. C {\bf 76},
065201 (2007).

\bm{sag07}
B. Saghai, J.-C. David, B. Julia-Diaz, and T.-S. H. Lee, 
Eur. Phys. J. A {\bf 31}, 512 (2007).

\bm{jul05}
B. Julia-Diaz, B. Saghai, F. Tabakin, W.-T. Chiang, T.-S. H. Lee, and
Z. Li, Nucl. Phys. {\bf A755}, 463c (2005). 
\bm{jul06}
B. Julia-Diaz, B. Saghai, T.-S. H. Lee, and F. Tabakin, Phys. Rev. C 
{\bf 73}, 055204 (2006).

\bm{chi01}
Wen-Tai Chiang, F. Tabakin, T.-S. H. Lee, and B. Saghai, Phys. Lett.
{\bf B517}, 101 (2001).

\bm{mar01a}
T. Mart, C. bennhold, H. Haberzettl, and L. Tiator, 
http://www.kph.uni-mainz.de/kaonmaid.html.

\bm{dav96}
J.~C. David, C. Fayard, G.-H. Lamot, B. Saghai, Phys. Rev. C {\bf 53},
2613 (1996).

\bm{mar01b}
T. Mart and C. Bennhold, Phys. Rev. C {\bf 61}, 012201(R) (1999).

\bm{mar06}
T. Mart and A. Sulaksono, Phys. Rev. C {\bf 74}, 055203 (2006);
P. Bydzovsky and T. Mart, Phys. Rev. C {\bf 76}, 065202 (2007).

\bm{ade85}
R.~A. Adelseck, C. Bennhold, and L.~E. Wright, Phys. Rev. C
{\bf 32}, 1681 (1985).

\bm{ade90}
R.~A. Adelseck, and B. Saghai, Phys. Rev. C {\bf 42}, 108 (1990).

\bm{wil91}
R.~A. Williams, C.~R. Ji, and S.~R. Cotanch, Phys. Rev. C {\bf 43},
452 (1991).

\bm{han01}
B.~S. Han, M.~K. Cheoun, K.~S. Kim, and I.~T. Cheon, Nucl. Phys.
{\bf A691}, 713 (2001).

\bm{li95}
Z. Li, Phys. Rev. C {\bf 52}, 1648 (1995).

\bm{lu95}
D. Lu, R. H. Landau, and S.~C. Phatak, Phys. Rev. C {\bf 52},
1662 (1995).

\bm{he08}
J.He, B. Saghai, Z. Li, Phys. Rev. C {\bf 78}, 035204 (2008).

\bm{bor07}
B. Borasoy, P. C. Bruns, U.-G. Meissner and R. Nissler, Eur. Phys. J.
A {\bf 34}, 161 (2007).

\bm{sch02}
O. Scholten, S. Kondratyuk, L. Van Daele, D. Van Neck, M. Waroquier and
A.Yu. Korchin, Acta Phys. Pol.{\bf B33}, 847 (2002).

\bm{new82}
R.G.~Newton,  \textit{Scattering theory of Waves and Particles}
  (Springer, New York, 1982).

\bm{kon00}
S. Kondratyuk and O. Scholten, Nucl. Phys. {\bf A677},396 (2000);
S. Kondratyuk and O. Scholten, Phys. Rev. C {\bf 62}, 025203 (2000).

\bm{kon02}
S. Kondratyuk and O. Scholten, Phys. Rev. C {\bf 65}, 038201;
{\it ibid.}  Phys. Rev. C {\bf 64}, 024005 (2001)

\bm{kor03}
A.Yu. Korchin and O.Scholten, Phys. Rev. C {\bf 68}, 045206 (2003).

\bm{sat96}
T. Sato and T.-S. H. Lee, Phys. Rev. C {\bf 54}, 2660 (1996)

\bm{chi04}
W. T. Chiang, B. Saghai, F. Tabakin, T.-S. H. Lee, Phys. Rev. C {\bf 69},
065208 (2004).

\bibitem{lut02}
M.F.M. Lutz and E.E. Kolomeitsev, Nucl. Phys. {\bf A700}, 193 (2002).

\bm{kon04}
S. Kondratyuk, K. Kubodera, F. Myhrer and O.Scholten, Nucl. Phys. {A736},
339 (2004).

\bm{pas00}
V. Pascalutsa, Nucl. Phys. {\bf A680}, 76 (2000).

\bm{pas01}
V. Pascalutsa, Phys. Lett. {\bf B503}, 85 (2001).

\bm{PDG}
W. M. Yao {\it et al.} ( Particle Data Group), J. Phys. G: Nucl. Part. Phys.
{\bf 33}, 1 (2006).

\bm{dav01}
R. M. Davidson and R. Workman, Phys. Rev. C {\bf 63}, 025210 (2001).

\bm{vir96}
Virginia Tech SAID data base, see the URL http://gwdac.phys.hwu.edu/, FA08
solution R. A.Arndt, I. I. Strakovsky, and R. L. Workman, Phys. Rev. C
{\bf 53}, 430 (1996); R. A. Arndt, W. J. Briscoe, I. I. Strakovsky and R. L.
Workman, Phys. Rev. C {\bf 74}, 045205 (2006).

\bm{kai97}
N. Kaiser, T. Waas and W. Weise, Nucl. Phys. {\bf A612}, 297 (1997).

\bm{sar05}
A. V. Sarantsev, V. A. Nikonov, A. V. Anisovich, E. Klempt, and U. Thoma,
Eur. Phys. J. A {\bf 25}, 441 (2005).

\bm{gui03}
M. Guidal, J. M. Laget, M. Vanderhaeghen, Nucl. Phys. {\bf A627},
645 (1997).

\bm{cor06}
T. Corthals, J. Ryckebusch, and T. Van Cauteren, Phys. Rev. C {\bf 73},
045207 (2006).

\bm{cor07a}
T. Corthals, T. Van Cauteren, J. Ryckebusch, and D. G. Ireland, Phys. Rev. C
{\bf 75}, 045204 (2007).

\bm{cor07b}
T. Corthals, T. Van Cauteren, P. Vancraeyveld, J. Ryckebusch, and D. G. 
Ireland, Phys. Lett. {\bf B 656}, 186 (2007).

\bm{van09}
P. Vancraeyveld, L. De Cruz, J. Ryckebusch, and T. Van Cauteren,
Phys. Lett. {\bf B681}, 428 (2009).

\bm{bjo64}
J. D. Bjorken and S. D. Drell, {\it Relativistic Quantum Mechanics}
(McGraw-Hill, New York, 1964).

\bm{cot04}
I.J. General and S.R. Cotanch, Phys. Rev. C {\bf 69}, 035202 (2004).

\bm{pas98}
V. Pascalutsa, Phys. Rev. D {\bf 58}, 096002 (1998).

\bm{luk06}
L. Jahnke and S. Leupold, Nucl. Phys. {\bf A778}, 63 (2006).

\end{thebibliography}
\end{document}